\documentclass[%
 preprintnumbers,
 amsmath,amssymb,
 aps,
]{revtex4-2}

\usepackage{graphicx}
\usepackage{dcolumn}
\usepackage{bm}

\usepackage{subcaption}

\begin{document}


\title{Graph Neural Network for Neutrino Physics Event Reconstruction}

\author{A. Aurisano}
\author{V Hewes}
\affiliation{University of Cincinnati, Cincinnati, OH 45221, USA}
\author{G. Cerati}
\author{J. Kowalkowski}
\affiliation{Fermi National Accelerator Laboratory, Batavia, IL 60510, USA}
\author{C. S. Lee}
\author{W. Liao}
\affiliation{Northwestern University, Evanston, Il 60208, USA}
\author{D. Grzenda}
\author{K. Gumpula}
\author{X. Zhang\footnote{also at {\it University of California, Los Angeles, Los Angeles, CA 90095,  USA}}}
\affiliation{Data Science Institute, University of Chicago, Chicago, IL 60637, USA}

\date{\today}

\begin{abstract}
Liquid Argon Time Projection Chamber (LArTPC) detector technology offers a wealth of high-resolution information on particle interactions, and leveraging that information to its full potential requires sophisticated automated reconstruction techniques. This article describes \textbf{NuGraph2}, a Graph Neural Network (GNN) for low-level reconstruction of simulated neutrino interactions in a LArTPC detector. Simulated neutrino interactions in the MicroBooNE detector geometry are described as heterogeneous graphs, with energy depositions on each detector plane forming nodes on planar subgraphs. The network utilizes a multi-head attention message-passing mechanism to perform background filtering and semantic labelling on these graph nodes, identifying those associated with the primary physics interaction with 98.0\% efficiency and labelling them according to particle type with 94.9\% efficiency. The network operates directly on detector observables across multiple 2D representations, but utilizes a 3D-context-aware mechanism to encourage consistency between these representations. Model inference takes 0.12~s/event on a CPU, and 0.005s/event batched on a GPU. This architecture is designed to be a general-purpose solution for particle reconstruction in neutrino physics, with the potential for deployment across a broad range of detector technologies, and offers a core convolution engine that can be leveraged for a variety of tasks beyond the two described in this article.

\end{abstract}

\maketitle

\section{\label{sec:intro}Introduction}

In recent years, machine learning (ML) applications have seen increasingly widespread use in neutrino physics, and more broadly in High Energy Physics (HEP). The NOvA experiment has utilized Convolutional Neural Networks (CNNs) for neutrino flavour identification \cite{Aurisano_2016}, prong identification~\cite{Psihas:2019ksa}, and cosmic filtering. CNN classification is currently used for the benchmark physics sensitivity at the next-generation Deep Underground Neutrino Experiment (DUNE) \cite{DUNE:2020gpm}; it has also been leveraged for lower-level reconstruction tasks \cite{DUNE:2022fiy}. MicroBooNE has been using CNNs for various tasks, including hit classification~\cite{MicroBooNE:2018kka} and particle identification~\cite{MicroBooNE:2020hho}.

However, dense CNN applications such as these have several limitations -- they typically operate on fixed-size inputs, which is often computationally inefficient for small events, and leads to the truncation of large events. Also, given the natural sparseness of most particle interactions, the dense representation leads to the majority of computations being wasted on empty space.

These limitations have been resolved through the adoption of sparse CNNs, utilizing tools such as MinkowskiEngine~\cite{choy20194d}. Sparse CNNs operate only on activated pixels within an infinite manifold, which resolves issues both with wasted computations and truncations. Applications in Liquid Argon Time Projection Chambers (LArTPCs) have shown great promise at particle hit identification and clustering~\cite{Koh:2020snv, MicroBooNE:2020yze, AbedAbud:2023cmz}.

Although sparse CNN applications resolve the major issues with dense CNNs, many more limitations inherent to the CNN paradigm persist. For instance, the kernel convolutions that define the approach require inputs to be fixed to a regular Euclidean grid; in the context of particle physics detectors, this can require arbitrary transformations to be applied to detector data. For instance, a 3D point cloud produced from detector hits must be aggregated into voxels in order to construct CNN inputs. Since the network output inhabits this voxelized domain, the end user must either perform this awkward transformation in reverse, or arbitrarily decide to work within the voxelized domain dictated by the choice of reconstruction algorithm.

This domain issue can be addressed with Graph Neural Networks (GNNs). Unlike CNNs, GNNs operate on a collection of graph nodes, along with edges that define their connectivity. This much more flexible data structure allows for network architectures that can operate on the data in its native domain without any transformations required. This flexibility also allows for more generic network architectures that are more agnostic to different outputs and detector technologies.

The HEPTrkX network architecture developed at the LHC served as a proof-of-concept for GNN applications in HEP. This model considered each energy deposition in the detector as a node in a graph, and predicted binary scores on graph edges between neighbouring detector layers to perform link prediction for track-forming~\cite{Farrell:2018cjr}. The continuation of this work, both within its successor collaboration, ExaTrkX~\cite{ExaTrkX:2021abe}, and elsewhere, explores more novel concepts such as hierarchical networks~\cite{Liu:2023siw}, equivariance~\cite{Murnane:2023kfm} and object condensation~\cite{Lieret:2023ydc}.

GNNs have also been explored within neutrino physics, albeit to a lesser extent. The IceCube collaboration explored a GNN architecture for high-energy $\nu_{\mu}$ selection, that considers each optical module in the detector as a graph node; this approach outperformed both traditional and CNN event selections~\cite{8614089}. Additionally, sparse CNN-based reconstruction techniques discussed previously have demonstrated the viability of utilizing GNNs at later stages to infer particle flow~\cite{Drielsma:2021jdv}.

This paper discusses NuGraph2, a GNN developed by the ExaTrkX collaboration for low-level particle reconstruction in neutrino physics. This network architecture conceptualizes the problem of particle reconstruction similarly to its HL-LHC counterpart, but arrives at a substantially different solution due to the domain differences between collider and neutrino physics.

NuGraph2 represents a substantial refinement of a proof-of-concept architecture~\cite{Hewes_2021}, retroactively named ``NuGraph1''. This earlier model was more directly inspired by the HEPTrkX architecture, performing link prediction between detector hits. This paradigm is effective at the LHC, where particle tracks are emitted radially outwards from the center of the detector, allowing tracks to be reconstructed by connecting hits on subsequent detector layers. However, in a neutrino physics context, where no such relationship between detector geometry and particle direction can be assumed, link prediction is no longer an effective method for hit clustering.

NuGraph2 operates by classifying hits on detector planes according to particle type, leveraging nexus connections between planes to utilize 3D context information. The network presented in this paper provides two outputs -- a binary filter that rejects hits that are not part of the primary physics interaction, and a semantic classifier that assigns each hit a particle type. However, the core message-passing engine discussed in this paper can be further leveraged for a broad range of applications such as event classification, clustering and more.

\section{Experimental setup} \label{sec:experiment}

\begin{figure}[htbp]
    \centering
    \includegraphics[width=0.7\textwidth]{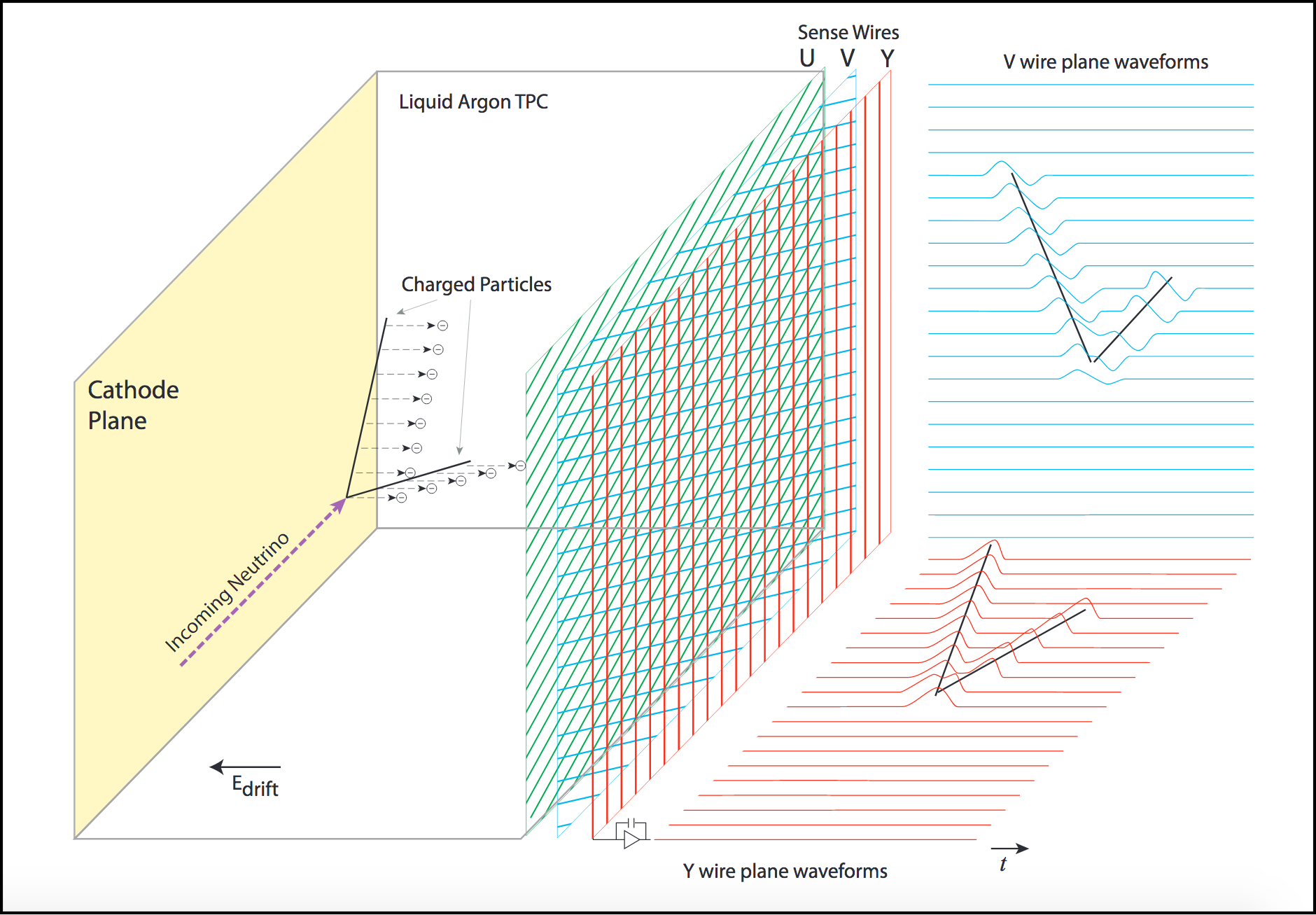}
    \caption{A diagram of the operation of LArTPC detector geometry.}
    \label{fig:detector}
\end{figure}

The results presented in this paper utilize an open dataset released by the MicroBooNE collaboration~\cite{Cerati:2023rtv}. This dataset consists of simulated $\nu_{e}$ and $\nu_{\mu}$ interactions in a realistic simulation of the MicroBooNE detector, with real data hits from cosmic interactions overlaid.

MicroBooNE is a LArTPC detector operating at Fermilab~\cite{MicroBooNE:2016pwy}, consisting of a cryostat containing cryogenically cooled liquid argon. Charged particles interacting in the argon volume produce ionization electrons that drift under an electric field to a series of three wire planes; these electrons induce electronic signals on the first two wire planes before being collected on the third. The three wire planes (U, V and Y) are oriented at an angle of $+60^{\circ}$, $-60^{\circ}$ and $0^{\circ}$ with respect to vertical, respectively. If the $t_{0}$ of a particle interaction is known, then the time at which signal pulses on TPC wires occur can be used as a measure of the signal's distance from the wires. Due to the differing wire angles, each TPC wire plane provides a unique 2D representation of a particle interaction; these three 2D representations can be combined to provide a full 3D reconstruction of a particle interaction.

In order to train the NuGraph2 architecture, this dataset is used to construct heterogeneous graph objects representing each neutrino interaction. For this paper, only hits passing the Pandora~\cite{MicroBooNE:2017xvs} neutrino slice selection are considered. This selection is based on 3D reconstruction and proximity and direction-based metrics, and separates 2D hits into slices representing distinct neutrino or cosmic ray interactions. Pulses on TPC wires are summarized into discrete Gaussian hits occurring on specific wires at specific times. An independent subgraph is formed for each wire plane, with the hits on a given plane used as the basis for that plane's graph nodes. Four input features are constructed for each graph node: the hit's wire index, time coordinate, and the integral and Root Mean Square (RMS) width of its Gaussian pulse.

Edges connecting hits within each detector plane's subgraph are formed using a Delaunay triangulation algorithm using each node's wire and time coordinates. Other edge formation algorithms, such as k-nearest-neighbour (kNN) and radius-based approaches, were also explored, but Delaunay triangulation was found to be both the most computationally efficient and the most performant. It generates a fully connected graph with a mixture of short-range and long-range connections, which allows it to maintain performance in the presence of spatially disconnected points due to detector effects such as dead wire regions and physics features such as disconnected electromagnetic shower segments.

In addition to planar subgraphs formed from detector hits, a ``nexus'' subgraph is constructed using spacepoints generated by SpacePointSolver~\cite{DUNE:2020ypp}, an upstream algorithm that groups together 2D hits across detector planes. This grouping of planar hits is utilized to define graph edges connecting planar nodes to nexus nodes. These nexus nodes are essentially ``virtual,'' in that they are defined solely by the graph edges connecting them to planar hits -- they have no dedicated input features, and no internal graph edges connecting nexus nodes to each other.

Planar graph nodes from the simulated neutrino interaction are assigned true semantic labels based on underlying simulation truth. The semantic labelling scheme considers five classes: \textbf{MIP}, which refers to particle tracks from minimum ionizing particles such as muons and charged pions; \textbf{HIP}, which refers to particle tracks from highly ionizing particles such as protons and charged kaons; \textbf{shower}, which refers to cone-shaped electromagnetic cascades produced by photons or electrons; \textbf{Michel}, which refers to Michel electrons produced at the end of a muon track; and \textbf{diffuse}, which refers to any particles with low hit multiplicity that do not produce an identifiable object, such as neutrons or electromagnetic activity due to Compton scattering.

The construction of graph objects from simulated events is achieved using a generalized workflow developed as part of this work. The NuML package~\footnote{\texttt{https://github.com/vhewes/numl}} is used to construct event-level HDF5 files containing tables summarizing low-level information such as simulated particles, simulated energy deposits and observed detector hits. These data structures are designed to be generalizable, and although the work described in this paper concerns the MicroBooNE detector, this workflow has also been successfully deployed in the context of other neutrino detectors, LArTPC and otherwise.

This event HDF5 format is then efficiently preprocessed into graph objects using the PyNuML package~\footnote{\texttt{https://github.com/vhewes/pynuml}}. This package leverages MPI parallelism to efficiently produce a large dataset into graph objects for model training at scale. Processed graphs are stored as individual datasets of compound data types in an HDF5 file. Using the compound data type allows us to store each graph sample in a contiguous space in the file, so that reading an entire graph can be completed in a single read call. Such design effectively minimizes the number of required I/O operations and thus reduces the overall I/O cost of the data loading phase.

PyNuML is designed to offer an efficient and flexible solution for many boilerplate ML tasks in HEP. It offers algorithms to efficiently generate ground truth labels from a tree of simulated particles, and then propagate them onto the observed detector hits, an approach which can be orders-of-magnitude more efficient than the typical ``backtracking'' approach taken by particle physicists. The highly-parallel event processing backbone is separated out from label generation and object construction algorithms, allowing the user to easily utilize, customize or override any aspect of the processing framework for their own needs.

\section{\label{sec:model}Model architecture}

\begin{figure}
    \centering
    \includegraphics[width=0.7\textwidth]{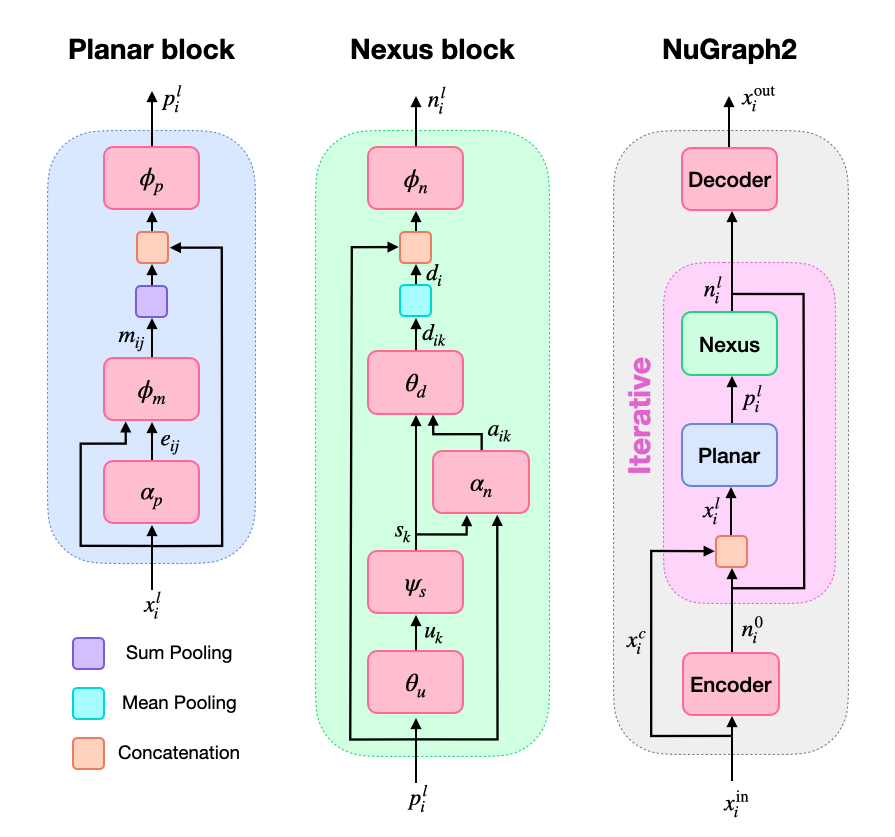}
    \caption{Block diagram for the planar (left) and nexus (center) blocks, and the overall structure of the NuGraph2 network architecture (right). $\phi$ denotes a planar convolution operator, $\psi$ a nexus convolution operator, $\theta$ a convolution that transforms between planar and nexus nodes, and $\alpha$ a categorical cross-attention convolution. The $i$ subscript refers to planar nodes, $j\in\mathcal{N}(i)$ their neighbouring planar nodes, and $k\in\mathcal{X}(i)$ the nexus nodes connected to each planar node; the $l$ superscript refers to the current iteration of message-passing.}
    \label{fig:architecture}
\end{figure}

The NuGraph2 network architecture consists of an encoder, an iterative message-passing engine, and a set of decoders; a high-level diagram of the network structure is provided in the right diagram of Fig.~\ref{fig:architecture}. Each iteration of the message-passing engine consists of two components -- a planar block responsible for propagating information internally within each detector plane, which is shown in the left diagram of Fig.~\ref{fig:architecture} and described in detail in Sec.~\ref{sec:planar-block}, and a nexus block responsible for passing information between detector planes, which is shown in the central diagram of Fig.~\ref{fig:architecture} and described in detail in Sec.~\ref{sec:nexus-block}.

All linear convolution layers are immediately followed by an activation function. The softmax and sigmoid functions are used for layers explicitly producing categorical and binary probabilities, respectively. for all other convolutional layers, a hyperbolic tangent activation function is used, as it was found to provide superior performance and stability when compared to the ReLU and Mish~\cite{misra2020mish} activation functions.

NuGraph2 utilizes a \textbf{categorical embedding} mechanism, in which the network learns a distinct set of hidden features for each particle category, as defined by hit semantic labels. Internal graph node feature tensors have shape $(N_{n}, N_{c}, N_{f})$, where $N_{n}$ is the number of graph nodes, $N_{c}$ is the number of semantic particle categories, and $N_{f}$ is the number of latent features per category.

Convolution layers operating on these categorical tensors utilize a \textbf{categorical linear} mechanism, in which a separate linear layer operates on the hidden features for each category, in a manner equivalent to grouped linear convolutions~\cite{imagenet}. A categorical linear convolution layer requires a factor of $N_{c}$ fewer neurons than an equivalent fully connected linear layer.

The result of this categorical convolution mechanism is that each category's hidden features are effectively ``unaware'' of the corresponding features for other particle categories, meaning that the network can learn information specific to each particle class.

We also utilize a specialized multi-head self-attention mechanism referred to as \textbf{categorical cross-attention}. This mechanism applies an MLP to a node feature tensor in a categorical embedding, convolving down to a single feature for each node category, and then applying a softmax across the categorical dimension. This produces a set of probabilities per category, which can then be applied to the original node feature tensor as self-attention weights.

When combined, the categorical linear and categorical cross-attention mechanisms provide an efficient and effective convolution mechanism. The application of categorical linear convolutions means that each category's features are allowed to evolve independently, while the categorical cross-attention mechanism allows for some controlled exchange of context information between categories. Categorical cross-attention is utilized to weight messages formed on graph edges during message-passing, suppressing information from categories that are not strongly activated while enhancing information from categories with strong activation.

Throughout this section, different types of convolution operator are denoted with different characters: $\phi$ for 2D planar convolutions, $\psi$ for 3D nexus convolutions, $\theta$ for convolutions that transform between planar and nexus nodes, and $\alpha$ for categorical cross-attention convolutions.

\subsection{Encoder}

NuGraph2's encoder serves to take the input graph node features $x^{\textrm{in}}$ and generate the initial categorical node embedding. First, the $(N_{n}, N_{i})$ input tensor is repeated $N_{c}$ times across a new second dimension to produce a new feature embedding on the planar nodes $x^{c}$ with shape $(N_{n},~N_{c},~N_{i})$, where $N_{i}$ is the number of node features. This is then used to generate the initial node embedding
\begin{equation}
    n^{0} = \phi_{e}(x^{c}),
\end{equation}
where $\phi_{e}$ is a categorical linear multi-layer perceptron (MLP) and $n^{0}$ is a planar node embedding with shape ($N_{n}$,~$N_{c}$,~$N_{f}$).

\subsection{Message passing}

After initial encoding to produce graph node features in a categorical embedding space, the network utilizes an iterative two-step message-passing mechanism as its core engine. This engine consist of a set of \textbf{planar blocks}, that operate independently on each 2D detector plane's subgraph, and a \textbf{nexus block} that exchanges information between 2D planes to allow for degeneracy breaking.

At the start of each message-passing iteration $l$, the planar node embedding $n_{i}^{l-1}$ is is concatenated with $x^{c}$ along the final dimension to form the planar node embedding $x_{i}^{l}$ with shape ($N_{n}$,~$N_{c}$,~$N_{f}+N_{i}$)

\subsubsection{Planar block} \label{sec:planar-block}

The \textbf{planar block} is responsible for propagating information internally inside each detector plane. The input features on each plane $x^{l}$ in a categorical embedding are convolved to produce a set of categorical cross-attention scores
\begin{equation}
    e_{ij} = \alpha_{p}(x^{l}_{i}, x^{l}_{j}),
    \label{eqn:planar-attention}
\end{equation}
where $\alpha_{p}$ is a categorical cross-attention block, and $j \in \mathcal{N}(i)$ are the neighbors of node $i$. The resulting attention weights are then used to weight messages formed on graph edges,
\begin{equation}
    m_{ij} = \phi_{m}(x_{i}, x_{j}, e_{ij}),
    \label{eqn:planar-message}
\end{equation}
where $\phi_{m}$ is a categorical linear MLP. These messages are then aggregated into graph nodes via sum pooling,
\begin{equation} \label{eqn:planar-aggr}
    m_{i} = \sum_{j \in \mathcal{N}(i)}m_{ij},
\end{equation}
and then concatenated with the input node features and convolved once more,
\begin{equation}
    p^{l}_{i} = \phi_{p}(m^{l}_{i}, x^{l}_{i}),
\end{equation}
where $\phi_{p}$ is a categorical linear MLP and $p_{i}^{l}$ is a planar node embedding with shape ($N_{n}$,~$N_{c}$,~$N_{f}$).

\subsubsection{Nexus block} \label{sec:nexus-block}

Following planar convolutions, the \textbf{nexus block} is responsible for mixing information between detector planes in order to break degeneracies inherent in each 2D representation. As an example, a long minimum ionising track (ie. a muon) propagating at a shallow angle with respect to the pitch of the wires on one detector plane can appear as a shorter, more highly ionising track (ie. a proton). The nexus block allows the network to cross-reference this representation with representations in other planes.

As discussed in Sec.~\ref{sec:experiment}, nexus nodes are ``virtual'' in the sense that they are defined solely by their connections to planar nodes. No message-passing is performed internally in the nexus space; instead, planar node features are propagated into the nexus nodes via message-passing, and the resulting nexus node features are convolved to form new nexus node features containing a mixture of information from all planes. These nexus features are then propagated back down to the planar nodes via message-passing, providing each planar subgraph with an injection of contextual information from other planes.

In order to accomplish this, node features in each planar subgraph are first used to form messages on the edges connecting the planar and nexus nodes
\begin{equation}
    u_{k} = \theta_{u}(p^{l}_{i}),
\end{equation}
where $k=\mathcal{X}(i)$ are the nexus nodes connected to planar node $i$, and $\theta_{u}$ is a simple message-passing convolution operator that transforms planar node features to nexus nodes. Since nexus nodes are derived from spacepoints that group together hits across detector planes, each nexus node is connected to a maximum of one planar node per plane, and so planar features are aggregated into nexus nodes via concatenation. In the event that a nexus node is not connected to any node on a given plane, that plane's contribution to the nexus node's features are set to zero.

Features on each nexus node are then passed forward into a categorical MLP to produce new nexus features
\begin{equation}
    s_{k} = \psi_{s}(u_{k}),
\end{equation}
where $\psi_{s}$ is a categorical linear MLP. Nexus node features are then passed back down to each planar subgraph, first by producing categorical cross-attention weights
\begin{equation}
    a_{ik} = \alpha_{n}(p^{l}_{i}, s_{k}),
\end{equation}
where $\alpha_{n}$ is a categorical cross-attention MLP, and then by using those weights to pass messages formed on edge features
\begin{equation}
    d_{ik} = \theta_{d}(s_{k}, a_{ik}),
\end{equation}
where $\theta_{d}$ is a categorical linear MLP. Messages are then aggregated into planar nodes using mean pooling,
\begin{equation}
    d_{i} = \frac{1}{|\mathcal{X}(i)|} \sum_{k \in \mathcal{X}(i)} d_{ik}.
\end{equation}
Here mean pooling is utilized instead of the sum pooling used in Eqn.~\ref{eqn:planar-aggr} for planar message-passing, as it yields stronger performance. These new planar node features are then shortcut connected with previous node features and convolved,
\begin{equation}
    n^{l}_{i} = \phi_{n}(p^{l}_{i}, d^{l}_{i}),
\end{equation}
where $\phi_{l}$ is a categorical linear MLP and $n_{i}^{l}$ is a planar node embedding with shape ($N_{n}$,~$N_{c}$,~$N_{f}$), which is then passed forward into either the subsequent iteration or the decoders.

\subsection{Decoders}

The output of the final message-passing iteration is passed into two decoders, which are responsible for producing the filter and semantic outputs. The filter decoder first flattens the categorical embedding tensor from a three-dimensional tensor with shape $(N_{n}, N_{c}, N_{f})$ to a two-dimensional tensor with shape $(N_{n}, N_{c} \times N_{f})$, and then convolves each nodes's features into a single binary classifier score by utilising a single linear convolutional layer followed by a sigmoid activation. The semantic decoder is similar in structure but retains the categorical embedding, convolving each category's features for each node down to a single score, and then taking the softmax activation across categories to produce categorical probability scores for each node.

\section{Training}

The graph dataset consists of 333771 events, which are subdivided according to a 90\%/5\%/5\% split into training, validation and testing samples containing 300395, 16688 and 16688 events respectively. At the start of each epoch, the training dataset is shuffled at random into 4693 batches of 64 events, with each of these batches forming the input to a single iteration of training.

NuGraph2 utilizes PyTorch~\cite{paszke2019pytorch} with PyG~\cite{fey2019fast} for our model, leveraging the model training infrastructure provided by PyTorch Lightning. Training is optimized using the AdamW optimizer~\cite{loshchilov2019decoupled} with a learning rate of 0.001, scheduled using the OneCycleLR scheduler~\cite{smith2018superconvergence} for a total of 80 epochs (375,440 iterations).

A simple binary cross-entropy is used to generate the loss for the hit filtering decoder. For the semantic decoder, we utilize the Recall Loss \cite{tian2022striking}, as it provides an optimal balance of precision and recall performance even in the presence of severe class imbalance. This approach was found to provide significantly more balanced performance than other techniques explored, such as the inverse frequency-weighted Categorical Cross-Entropy Loss, which naively optimized recall for minority classes at the expense of precision, and the Focal Loss~\cite{lin2017focal}, which failed to learn on minority classes altogether.

Loss functions for multi-task training are balanced by utilising a temperature mechanism to automatically find an optimal performance balance between both tasks~\cite{kendall2018multitask}.

\section{Results} \label{sec:results}

\subsection{Model performance} \label{sec:performance}

Network performance is quantified in terms of \textbf{recall} (efficiency) and \textbf{precision} (purity), which are defined as

\begin{align}
    \textrm{Recall} &= \frac{\textrm{TP}}{\textrm{TP} + \textrm{FN}} & \textrm{Precision} &=  \frac{\textrm{TP}}{\textrm{TP} + \textrm{FP}}
\end{align}
where TP, FP and FN represent true positives, false positives and false negatives, respectively. We also utilize the \textbf{receiver-operator characteristic} (ROC) curve, in which the true positive rate and false positive rate are computed and visualized for a range of probability score thresholds.

\begin{figure}
    \centering
    \includegraphics[width=0.7\linewidth]{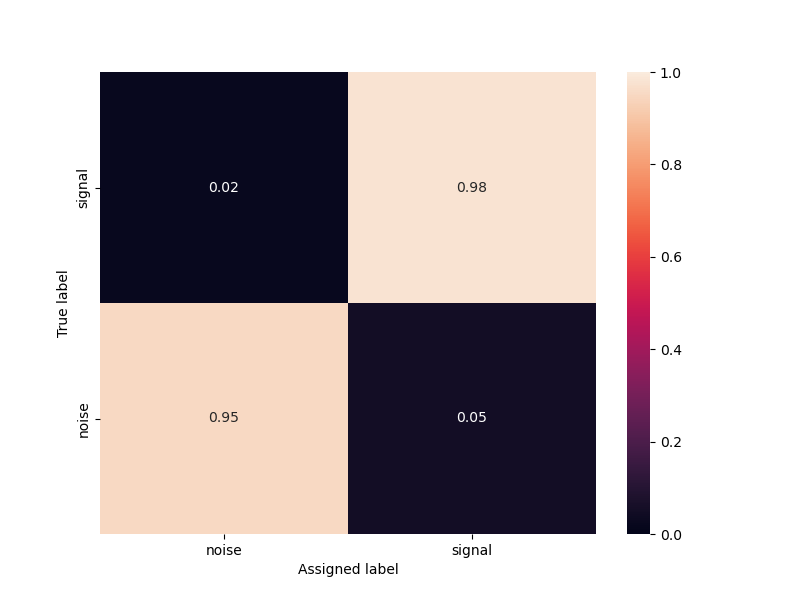}
    \caption{Confusion matrix for hit filtering, with probabilities normalized per predicted class in order to highlight recall.}
    \label{fig:filter-confusion}
\end{figure}

\begin{figure}
    \centering
    \begin{subfigure}{0.45\linewidth}
        \includegraphics[width=\linewidth]{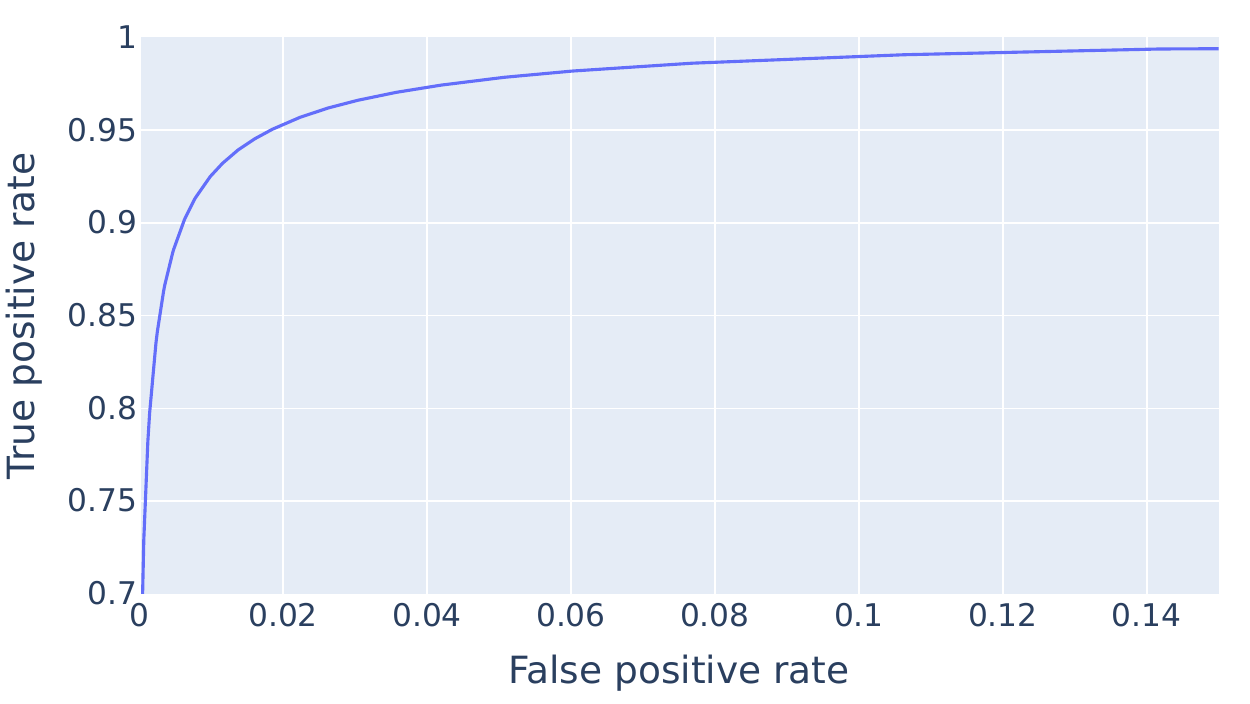}
        \caption{ROC curve for filter prediction.}
        \label{fig:roc-filter}
    \end{subfigure}
    \begin{subfigure}{0.45\linewidth}
        \includegraphics[width=\linewidth]{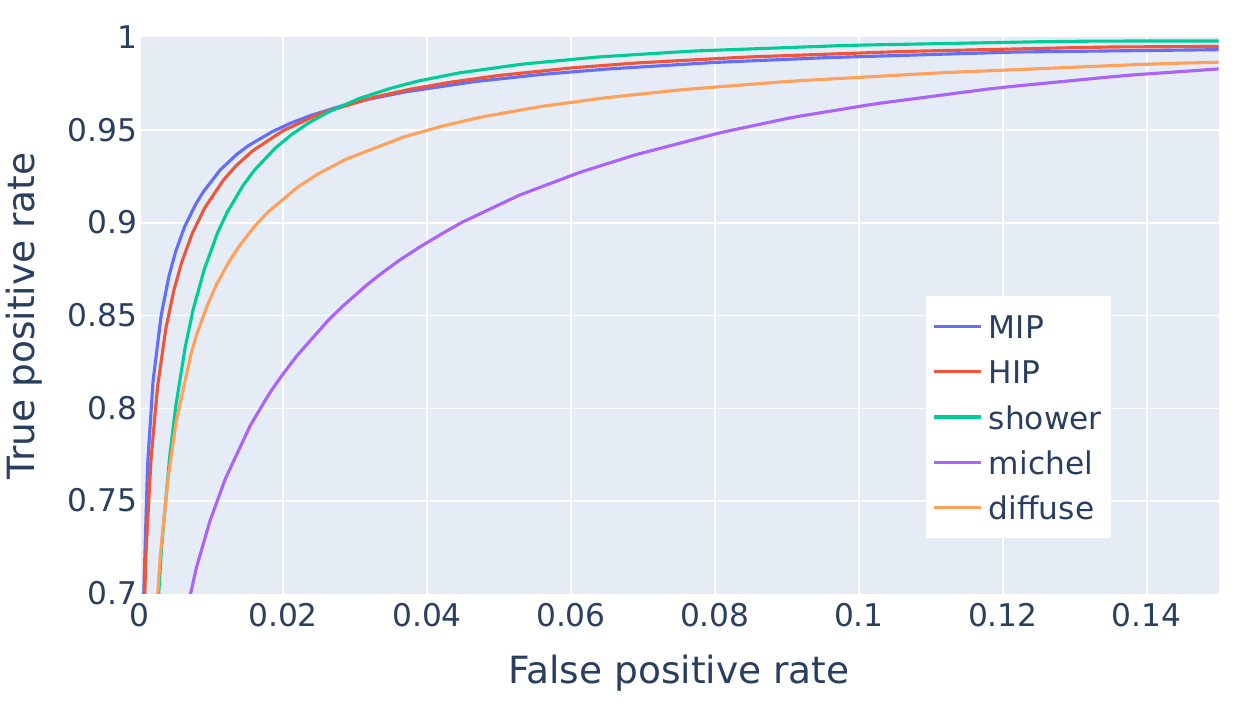}
        \caption{ROC curves for each semantic class.}
        \label{fig:roc-semantic}
    \end{subfigure}
    \caption{Receiver-Operator Characteristic (ROC) curves for model predictions.}
    \label{fig:roc}
\end{figure}

For the network's binary filter output, positive and negative predictions are determined per hit by applying a probability threshold of 0.5. The filter decoder achieves a recall of 0.980 and a precision of 0.979; a recall confusion matrix is shown in Fig.~\ref{fig:filter-confusion}. Due to the preponderance of signal hits over background hits in the training set, the recall metric for signal hits is larger than for background hits when applying a balanced threshold of 0.5. The filter output can be optimized towards selecting signal hits or rejecting background hits by applying a different threshold, as reflected in the filter ROC curve shown in Fig.~\ref{fig:roc-filter}.

\begin{figure}
    \centering
    \begin{subfigure}{\textwidth}
        \includegraphics[height=0.45\textheight]{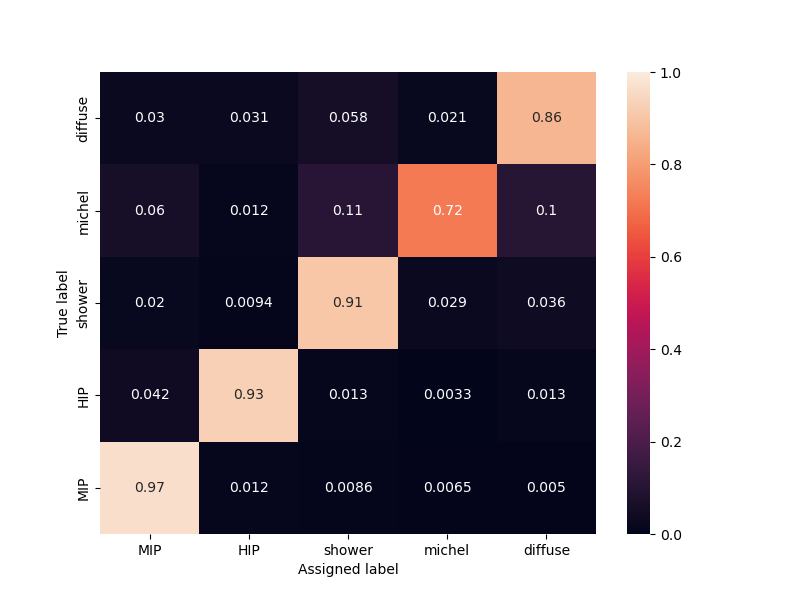}
        \caption{Confusion matrix normalized by true semantic class.}
        \label{fig:semantic-confusion-true}
    \end{subfigure}
    \begin{subfigure}{\textwidth}
        \includegraphics[height=0.45\textheight]{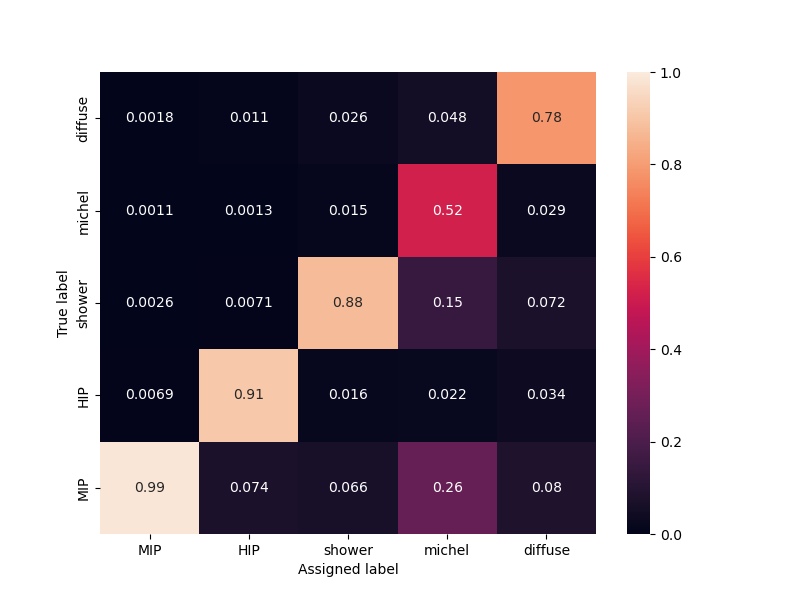}
        \caption{Confusion matrix normalized by predicted semantic class.}
        \label{fig:semantic-confusion-pred}
    \end{subfigure}
    \caption{Confusion matrices for hit semantic segmentation.}
    \label{fig:semantic-confusion}
\end{figure}

For the network's semantic probability outputs (using the labelling scheme defined in Sec.~\ref{sec:experiment}), a discrete class prediction per hit is determined by selecting the class with the largest probability score. The semantic decoder achieves an overall recall and precision of 0.949; these values are equal due to the scheme used to weight the loss function. Recall and precision confusion matrices are shown in Fig.~\ref{fig:semantic-confusion}.

The semantic predictions are most performant for the MIP class, which is the most highly-represented class in the training dataset. This is due to the abundance of muons in the training set, coupled with the fact that muons are minimum-ionizing particles that produce long tracks, and therefore many detector hits. Conversely, the semantic predictions are least accurate for the Michel class, which is the least represented class in the dataset, due to a relative scarcity and low hit multiplicity.

The most common modes of semantic misclassification are consistent with qualitative expectations. The HIP and MIP classes are topologically similar, as they both form particle tracks, and they are typically distinguished through differences in track length and energy deposition. One of the network's primary failure modes is the misclassification of HIP hits as MIP-like -- 4.7\% of true HIP hits are predicted to be MIP-like.

The shower, Michel and diffuse classes have similar conceptual overlap, as all three more frequently consist of smaller energy depositions -- electromagnetic showers are typically formed from a cascade of smaller and frequently disconnected energy depositions, while Michel electrons are short tracks, and diffuse activity often manifests as small and disconnected energy depositions. This is reflected in the network output, as 11\% and 10\% of true Michel hits are misclassified as shower-like and diffuse respectively. Similarly, 5.8\% and 2.1\% of true diffuse hits are misclassified as shower-like and Michel-like, respectively.

Another primary failure mode is between the MIP and Michel classes, which are strongly correlated as Michel electrons always occur at the end of a muon track. This confusion primarily manifests as true MIP hits being misclassified as Michel-like, which is a consequence of the weighting scheme applied to the loss function. True MIP hits are misclassified as Michel-like at a rate of only 0.65\%, but due to class imbalance in the dataset, these misclassified hits constitute 26\% of hits predicted to be Michel-like by the network. By contrast, only 6\% of Michel hits are misclassified as MIP-like, and these hits only constitute 0.1\% of predicted MIP hits.

\begin{figure}
    \centering
    \begin{subfigure}{0.45\textwidth}
        \includegraphics[width=\textwidth]{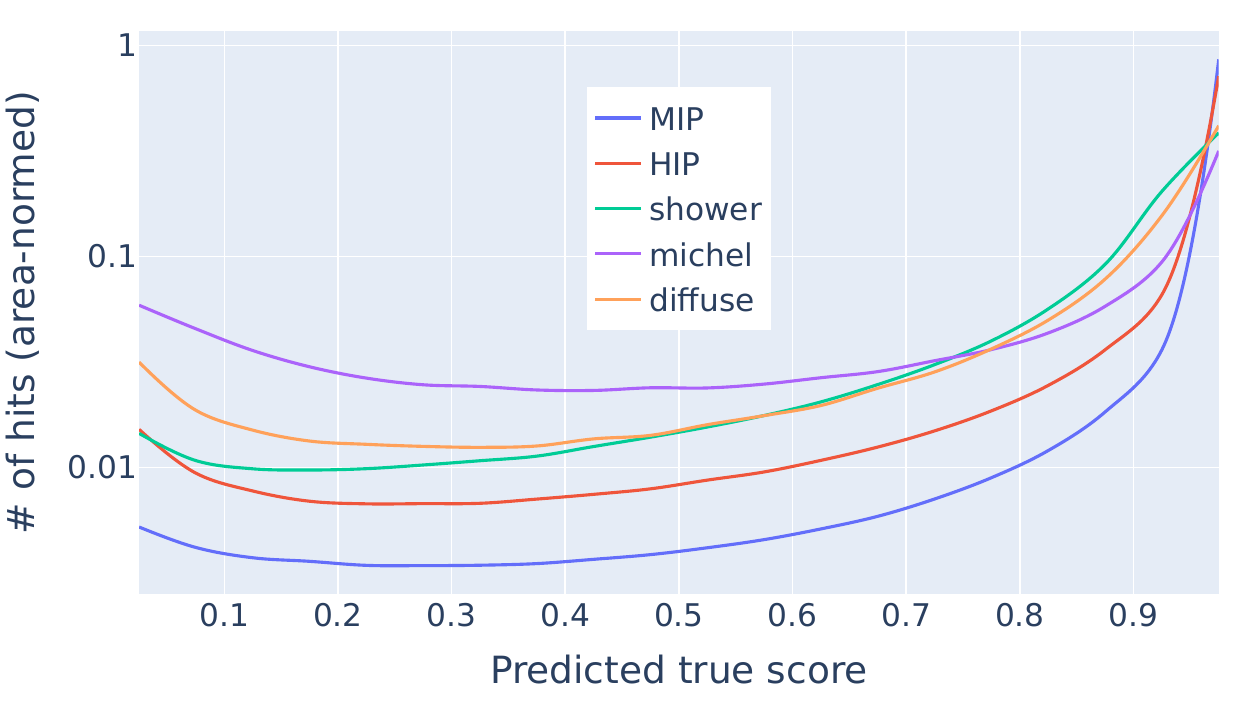}
        \caption{True score distributions}
        \label{fig:score-true}
    \end{subfigure}
    \begin{subfigure}{0.45\textwidth}
        \includegraphics[width=\textwidth]{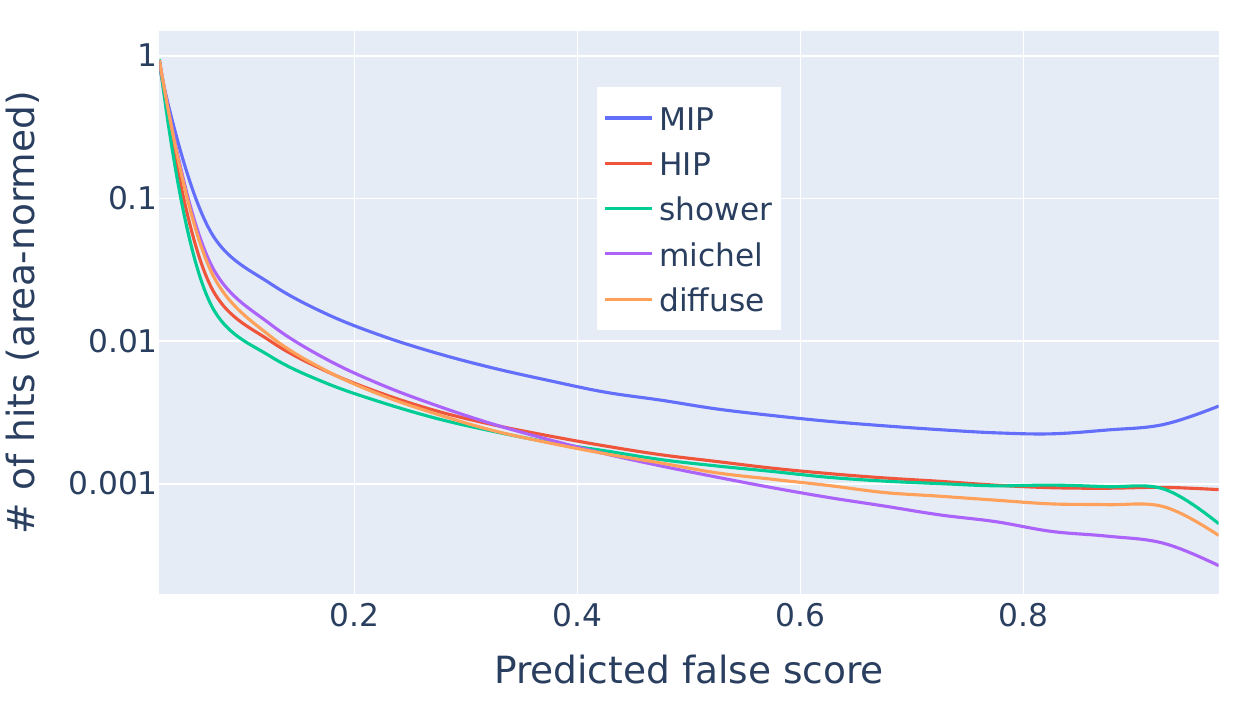}
        \caption{False score distributions}
        \label{fig:score-false}
    \end{subfigure}
    \caption{Predicted score distributions for each semantic category, separated into detector hits that belong to that category in truth (left), and hits that belong to other categories in truth (right). These distributions are area-normalized so their shapes can be compared.}
    \label{fig:score}
\end{figure}

Predicted score distributions for each semantic class are shown in Fig.~\ref{fig:score}, while ROC curves for each class are shown in Fig.~\ref{fig:roc-semantic}.

For semantic hit labelling, we also define a metric \textbf{consistency} as the proportion of nexus nodes for which all connected planar nodes share the same semantic label, in order to quantify the compatibility of semantic classification across different detector planes. Semantic labels in ground truth have a consistency of 98.04\%, while semantic labels predicted by NuGraph2 have a consistency of 94.81\%. By contrast, a planar-only model that considers each detector plane as an independent graph by omitting the nexus phase of message-passing achieves a consistency of only 67.38\%, demonstrating the nexus block's role in producing a coherent representation across detector planes.

\subsection{Inference benchmarks} \label{sec:inference}

\begin{figure}
    \centering
    \includegraphics[width=0.6\textwidth]{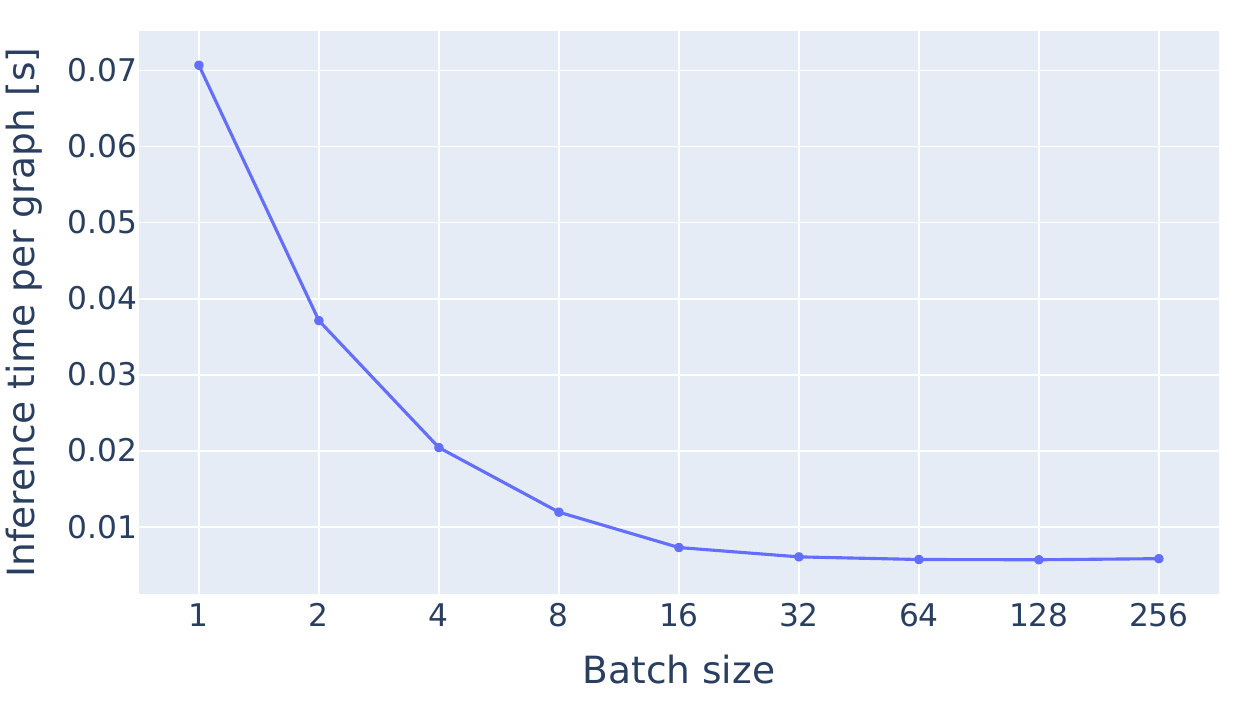}
    \caption{Inference time per graph on GPU as a function of batch size.}
    \label{fig:inference-time}
\end{figure}

Inference benchmarks were carried out on an Nvidia DGX GPU cluster. Evaluating NuGraph2 model output event-by-event on CPU takes 0.12~s/event. Inference on GPU reaches 0.005~s/event for sufficiently large batch sizes; GPU inference time as a function of batch size is shown in Fig.~\ref{fig:inference-time}.

NuGraph2 inference has also been implemented in the LArSoft software framework~\cite{Snider:2017wjd}, using a TorchScript-compiled model running on CPU. This preliminary implementation achieves an inference time of 0.3~s/event, which accounts for both graph construction and model inference time. Peak CPU memory utilization is currently 3.1~GB; memory overhead can be further improved, but this work is beyond the scope of this paper. A mature and flexible inference workflow that utilizes the SONIC interface~\cite{Cai:2023ldc} to run graph construction and model inference natively in Python is currently in development.

\subsection{Hyperparameter optimization} \label{sec:optimization}

\begin{figure}
    \centering
    \begin{subfigure}{0.45\textwidth}
        \includegraphics[width=\textwidth]{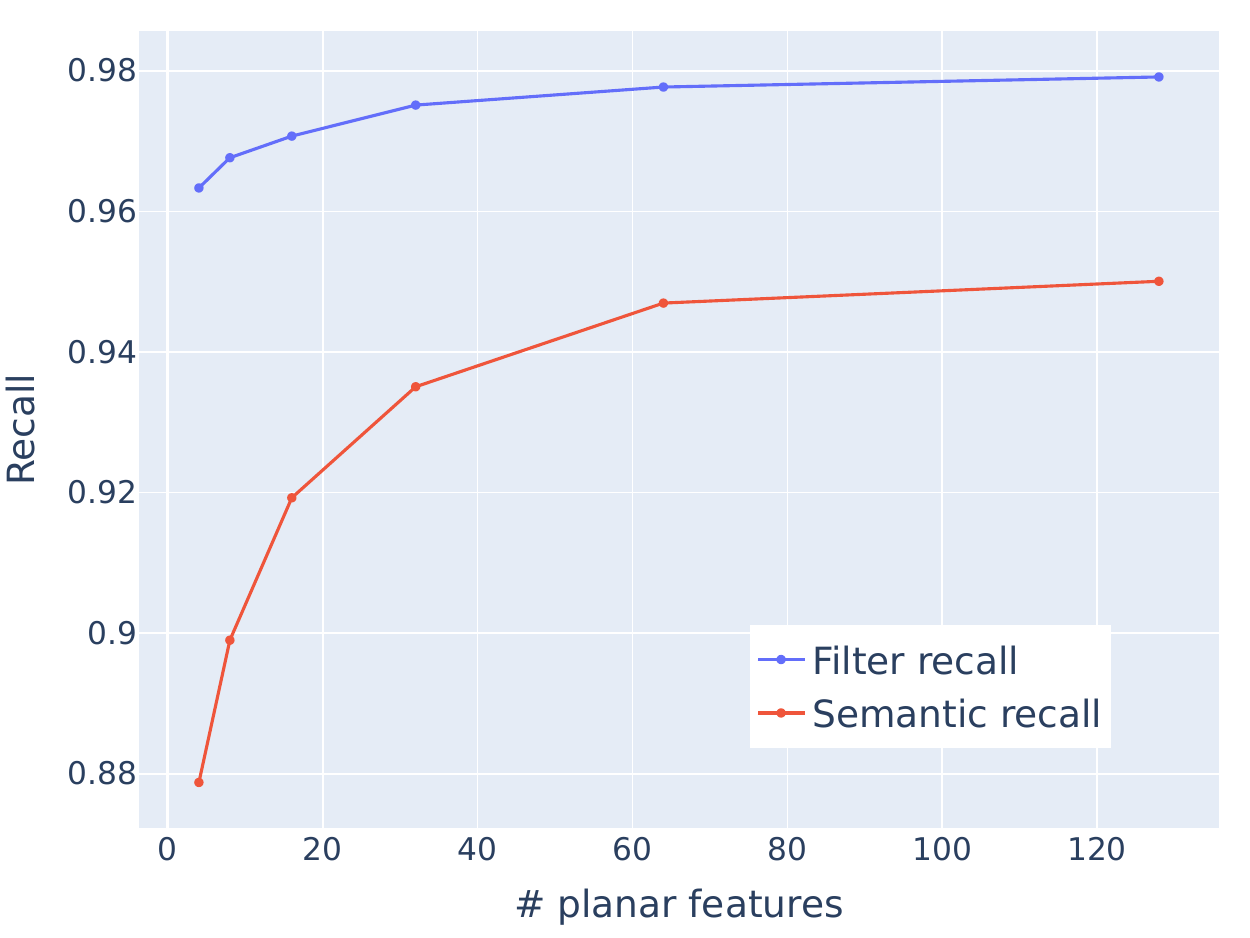}
        \caption{Network performance as a function of number of planar features}
        \label{fig:optimization-planar}
    \end{subfigure}
    \begin{subfigure}{0.45\textwidth}
        \includegraphics[width=\textwidth]{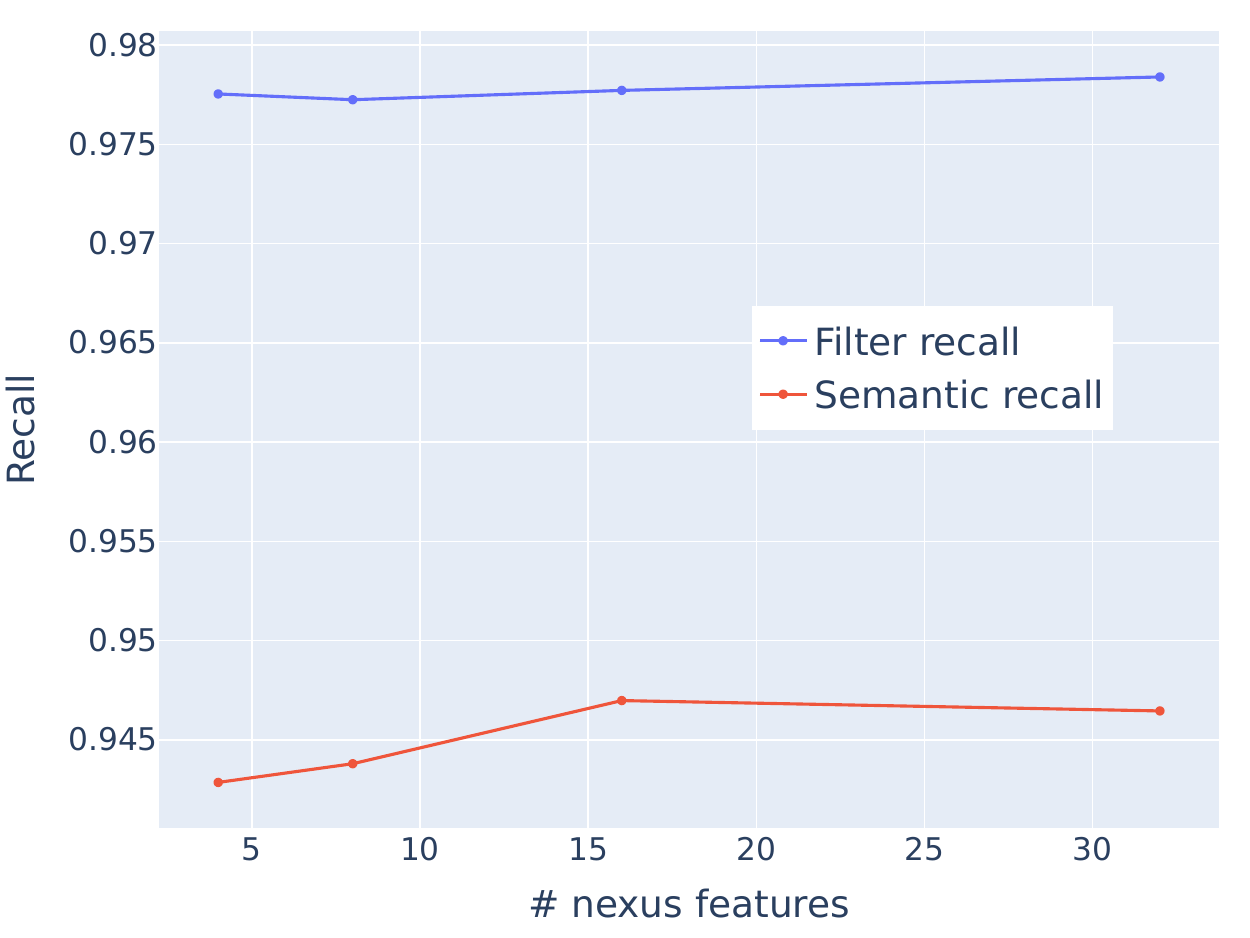}
        \caption{Network performance as a function of number of nexus features}
        \label{fig:optimization-nexus}
    \end{subfigure}
    \caption{Network performance as a function of number of hidden features utilized.}
    \label{fig:optimization}
\end{figure}

\begin{table}
    \centering
    \begin{tabular}{c|c|c|c|c|c}
         Variant & Planar feats. & Nexus feats. & Filter recall & Semantic recall  & Train GPU mem. (GB)\\
         \hline 
         nominal & 64 & 16 & 0.978 & 0.947 & 6.19 \\
         planar-4 & 4 & 16 & 0.963 & 0.879 & 0.69 \\
         planar-8 & 8 & 16 & 0.968 & 0.899 & 1.12 \\
         planar-16 & 16 & 16 & 0.971 & 0.919 & 1.83 \\
         planar-32 & 32 & 16 & 0.975 & 0.935 & 3.29 \\
         planar-128 & 128 & 16 & 0.979 & 0.950 & 11.98 \\
         nexus-4 & 64 & 4 & 0.978 & 0.943 & 6.18 \\
         nexus-8 & 64 & 8 & 0.977 & 0.944 & 6.19 \\
         nexus-32 & 64 & 32 & 0.978 & 0.946 & 6.19 \\
    \end{tabular}
    \caption{Network performance and efficiency metrics for different model configurations.}
    \label{tab:optimization}
\end{table}

The number of hidden parameters in each NuGraph2 convolution layer are configurable via hyperparameters, implemented independently for planar convolutions and nexus convolutions. The default values were chosen for a balance of performance and memory overhead. Network performance as a function of the number of node features and nexus features is shown in Fig.~\ref{fig:optimization}. Network performance is primarily dependent on the size of the planar embedding, with a weaker dependence on the size of the nexus embedding. The nominal sizes of 64 and 16 features for the planar and nexus embeddings, respectively, were chosen as larger embeddings incur significant memory overhead costs while network performance plateaus. More details on all model variants are provided in Tab.~\ref{tab:optimization}.

\subsection{Event displays} \label{sec:event-displays}

In order to provide context for the model performance metrics described above, this section discusses NuGraph2's performance on a series of representative events, designed to highlight events where the network performs well, and also events that highlight common failure modes. All events shown are drawn from the test partition of the dataset, and so were not used during network training.

For the purposes of this section, the network's filter prediction is converted into a binary score by applying a threshold at 0.5. This threshold is configurable, and can be varied to optimize towards signal selection or background rejection. Similarly, the network's classwise semantic probabilities are reduced to a discrete semantic label for each hit by selecting the class with the highest probability.

\subsubsection{Muon interaction}

\begin{figure}
    \centering
    \begin{subfigure}{0.45\textwidth}
        \includegraphics[width=\textwidth]{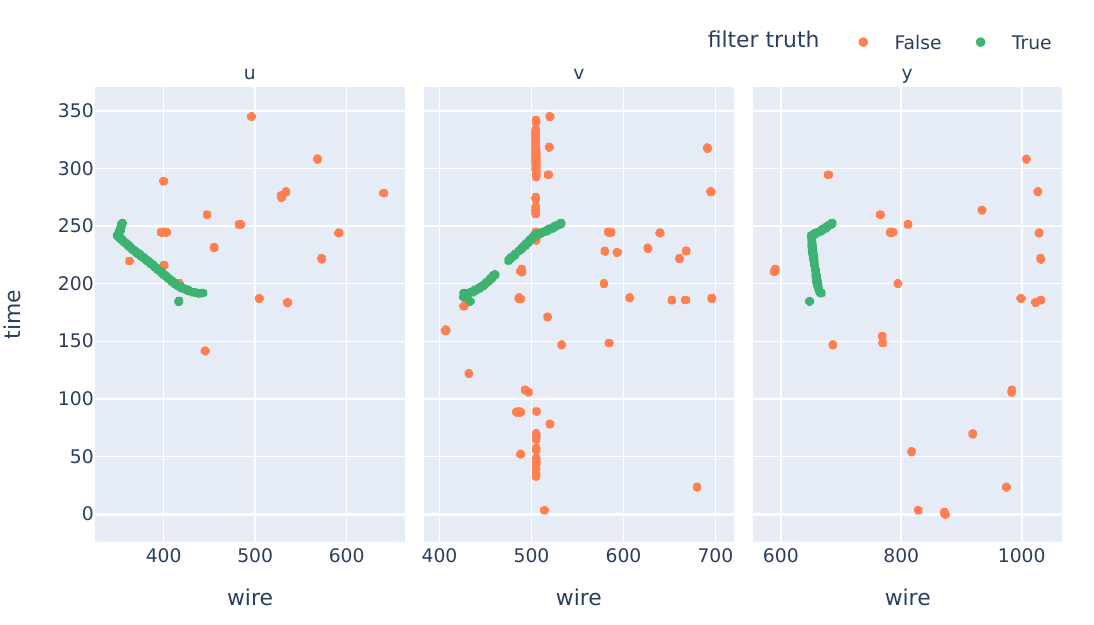}
        \caption{Filter truth}
    \end{subfigure}
    \begin{subfigure}{0.45\textwidth}
        \includegraphics[width=\textwidth]{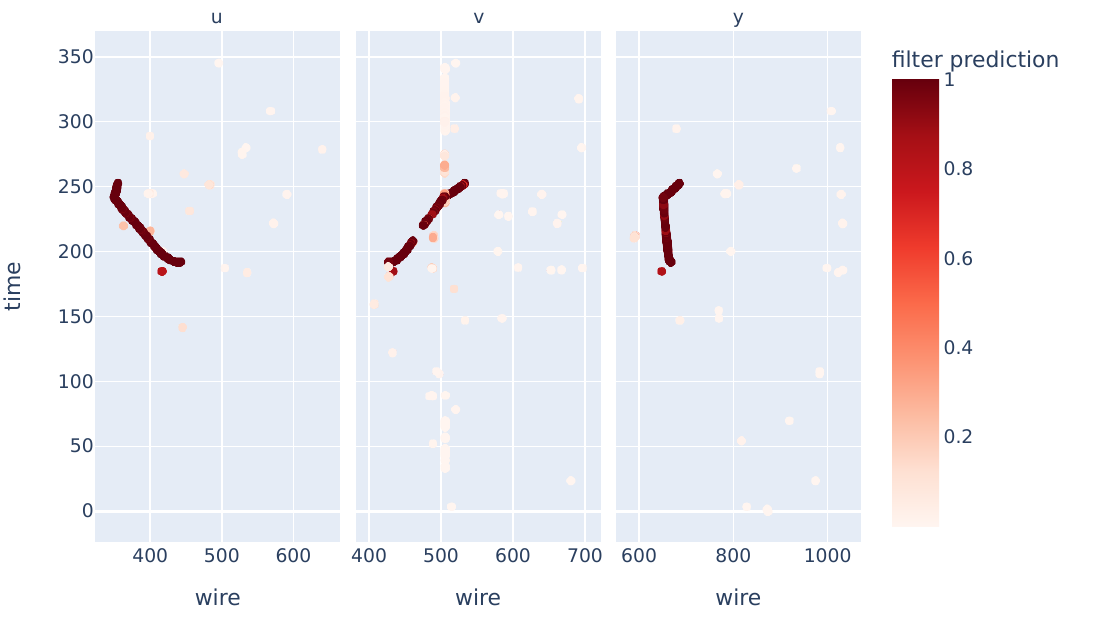}
        \caption{Filter prediction}
    \end{subfigure}
    \begin{subfigure}{0.45\textwidth}
        \includegraphics[width=\textwidth]{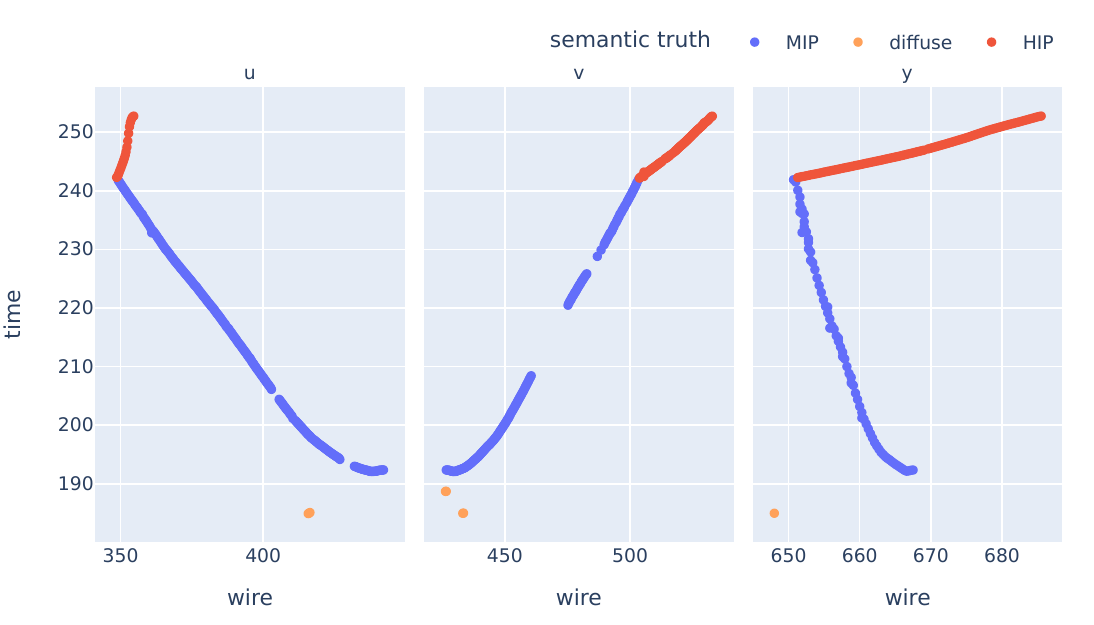}
        \caption{Semantic truth, filtered by truth}
    \end{subfigure}
    \begin{subfigure}{0.45\textwidth}
        \includegraphics[width=\textwidth]{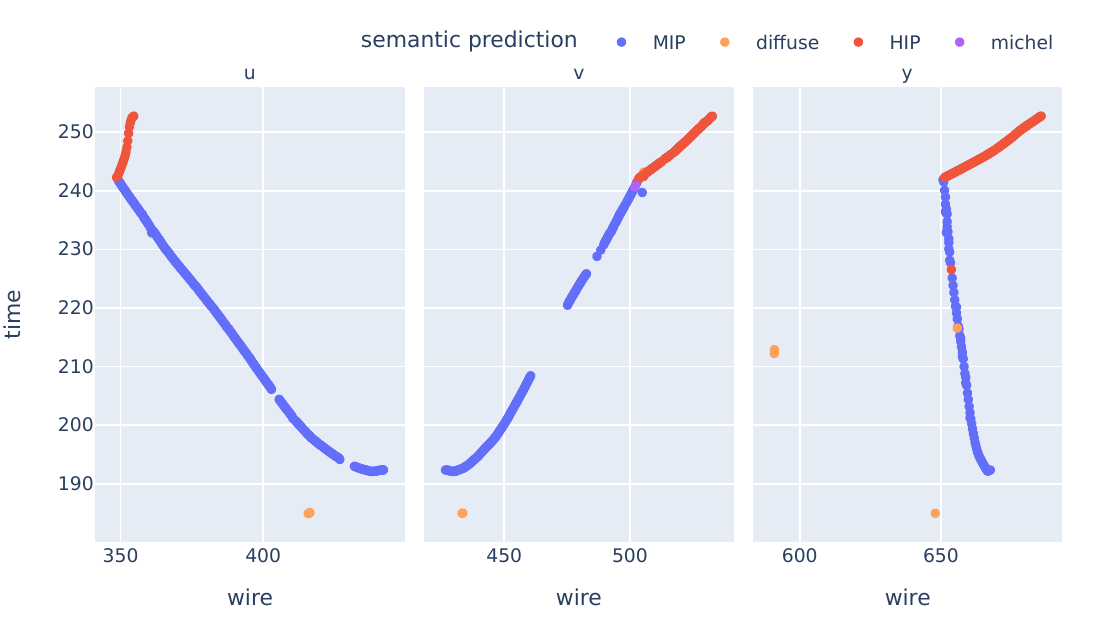}
        \caption{Semantic prediction, filtered by prediction}
    \end{subfigure}
    \caption{Event displays for the first representative event: Run 7049, Subrun 391, Event 19599 of the MicroBooNE open data release.}
    \label{fig:evd-1}
\end{figure}

The first representative event is Run 7049, Subrun 391, Event 19599 of the MicroBooNE open data release, shown in Fig.~\ref{fig:evd-1}. This event is a charged current quasielastic (CCQE) $\nu_{\mu}$ interaction, consisting of a longer track produced by a muon with 378~MeV momentum, and a shorter track produced by a proton with 740~MeV momentum. This event was selected as an example of a topologically simple event that is well-reconstructed by the network.

This event consists of 1058 hits across all wire planes, of which 777 are hits from the simulated neutrino interaction. The filter decoder correctly classifies 1053 hits; of the 5 misclassified hits, 2 are signal hits misclassified as background, and 3 are background hits misclassified as signal. The semantic decoder misclassifies 6 of the 777 signal hits -- 5 MIP hits are misclassified (2 as HIP, 2 as Michel and 1 as diffuse), and a single HIP hit is misclassified as diffuse.

This neutrino interaction is a clean two-prong topology, and the network is able to reject background hits and classify signal hits with $>99\%$ efficiency.

\subsubsection{Shower interaction}

\begin{figure}
    \centering
    \begin{subfigure}{0.45\textwidth}
        \includegraphics[width=\textwidth]{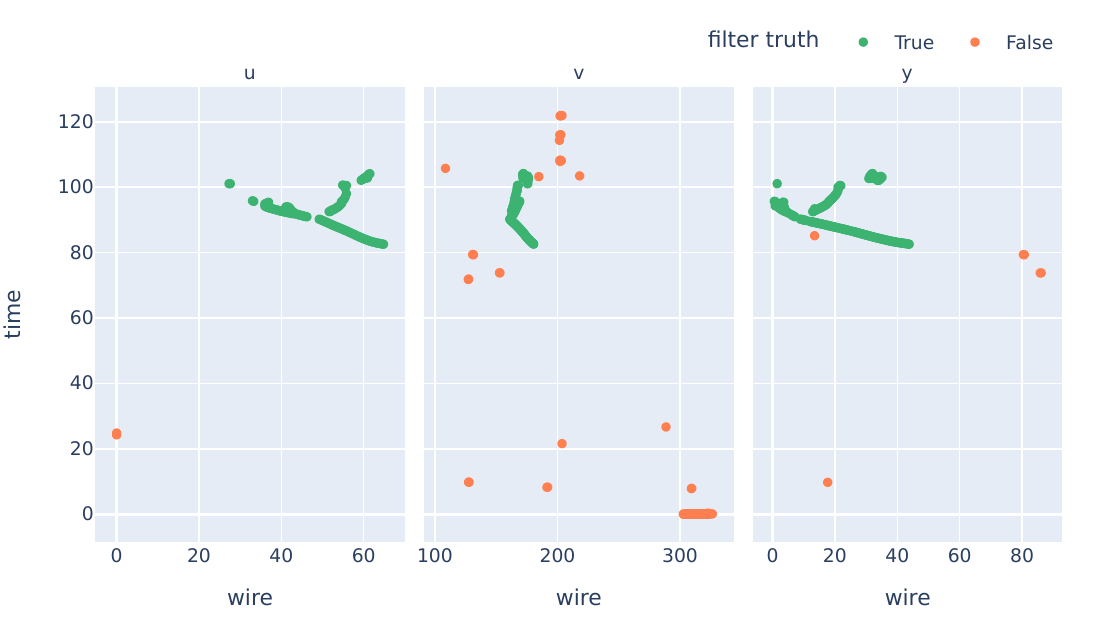}
        \caption{Filter truth}
    \end{subfigure}
    \begin{subfigure}{0.45\textwidth}
        \includegraphics[width=\textwidth]{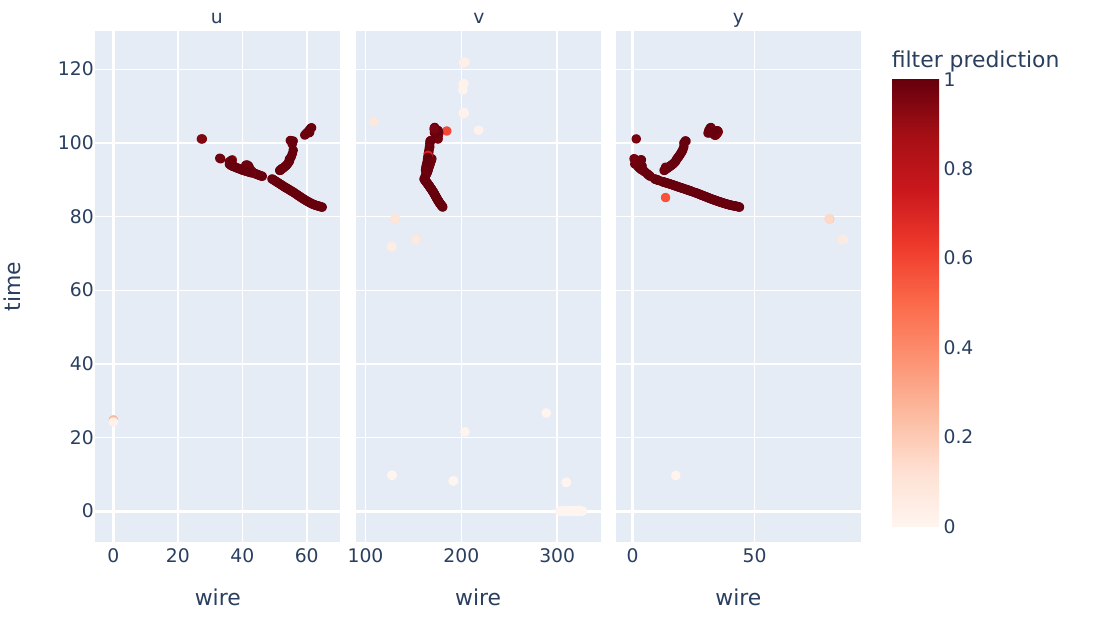}
        \caption{Filter prediction}
    \end{subfigure}
    \begin{subfigure}{0.45\textwidth}
        \includegraphics[width=\textwidth]{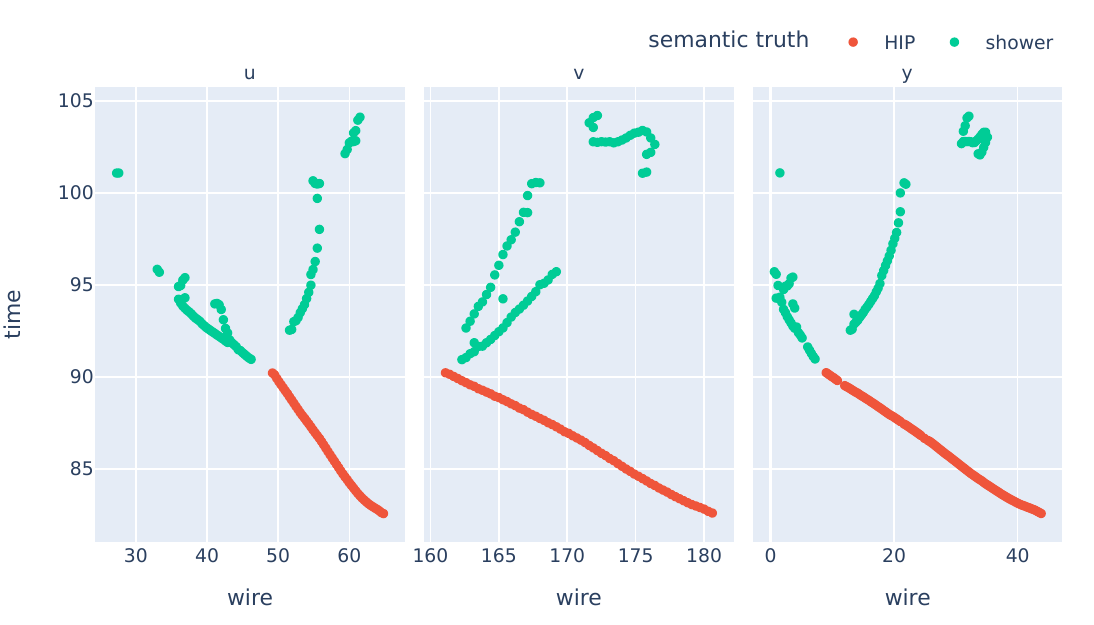}
        \caption{Semantic truth, filtered by truth}
    \end{subfigure}
    \begin{subfigure}{0.45\textwidth}
        \includegraphics[width=\textwidth]{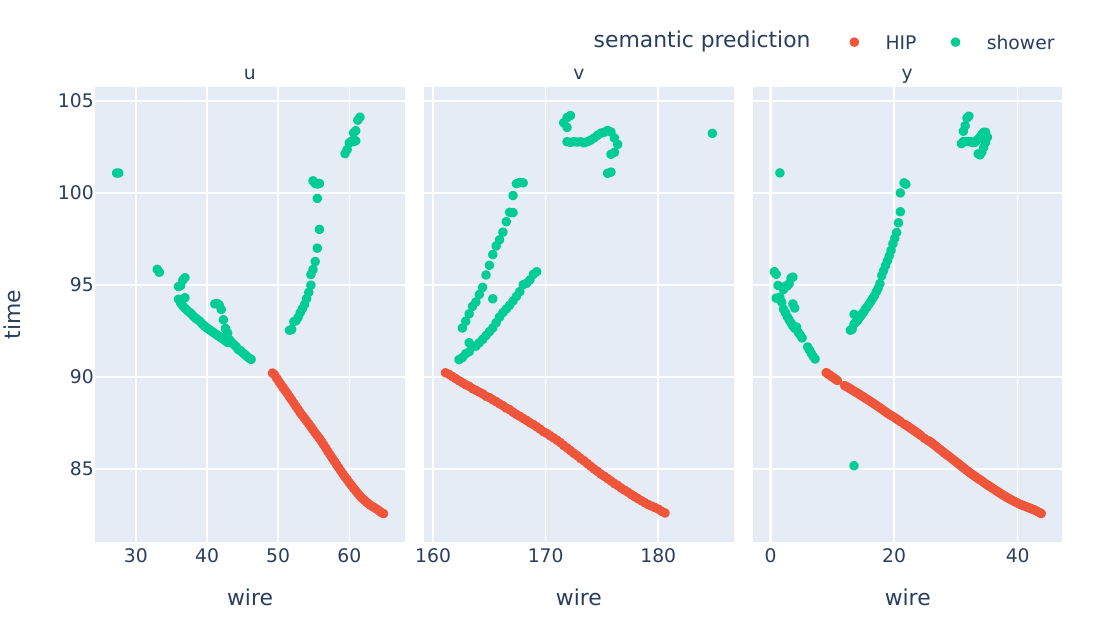}
        \caption{Semantic prediction, filtered by prediction}
    \end{subfigure}
    \caption{Event displays for the second representative event: Run 6999, Subrun 11, Event 595 of the MicroBooNE open data release.}
    \label{fig:evd-2}
\end{figure}

The second representative event is Run 6999, Subrun 11, Event 595 of the MicroBooNE open data release, shown in Fig.~\ref{fig:evd-2}. This event is a neutral current (NC) interaction, consisting of a track produced by a proton with 719~MeV momentum and a $\pi^{0}$ meson with 91~MeV momentum that is not observed directly, but decays into a pair of photons with momenta of 60~MeV and 103~MeV, each of which produces an electromagnetic (EM) shower. This event was selected as an example of a topologically simple event containing EM showers that is well-reconstructed by the network.

This event consists of 568 hits across all wire planes, of which 469 are hits from the simulated neutrino interaction. The filter decoder correctly classifies 566 hits, and misclassifies 2 background hits as signal. The semantic decoder correctly classifies all signal hits.

This event is easy to reconstruct, as background hits are largely disconnected from the physics interaction, and the photon showers are offset from the neutrino vertex. Under these conditions, the network is able to achieve near-perfect performance.

\subsubsection{Michel electron}

\begin{figure}
    \centering
    \begin{subfigure}{0.45\textwidth}
        \includegraphics[width=\textwidth]{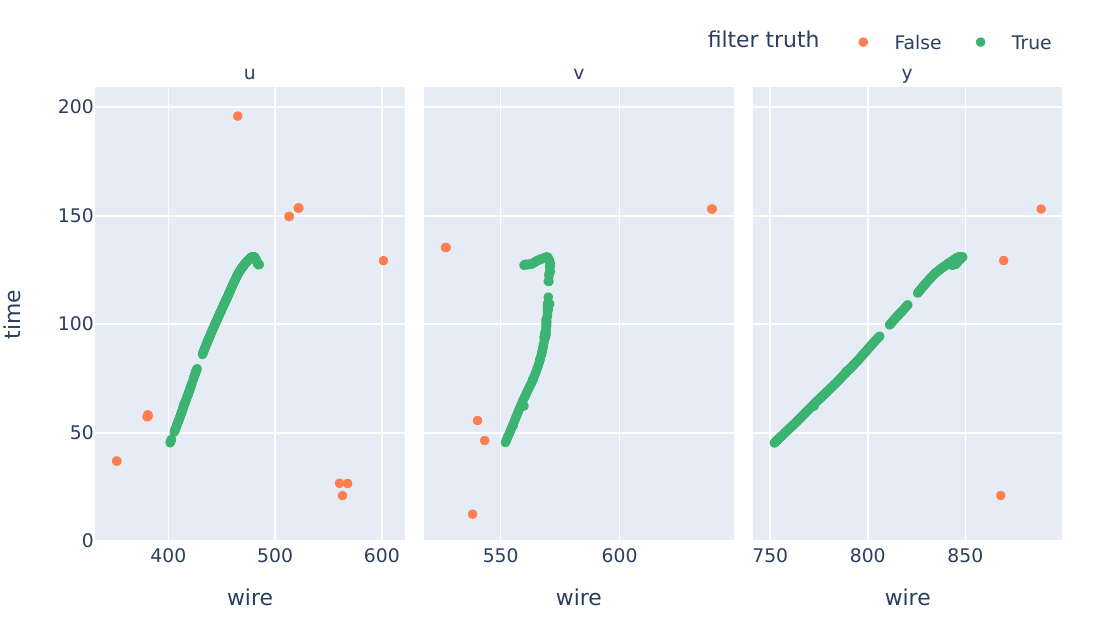}
        \caption{Filter truth}
    \end{subfigure}
    \begin{subfigure}{0.45\textwidth}
        \includegraphics[width=\textwidth]{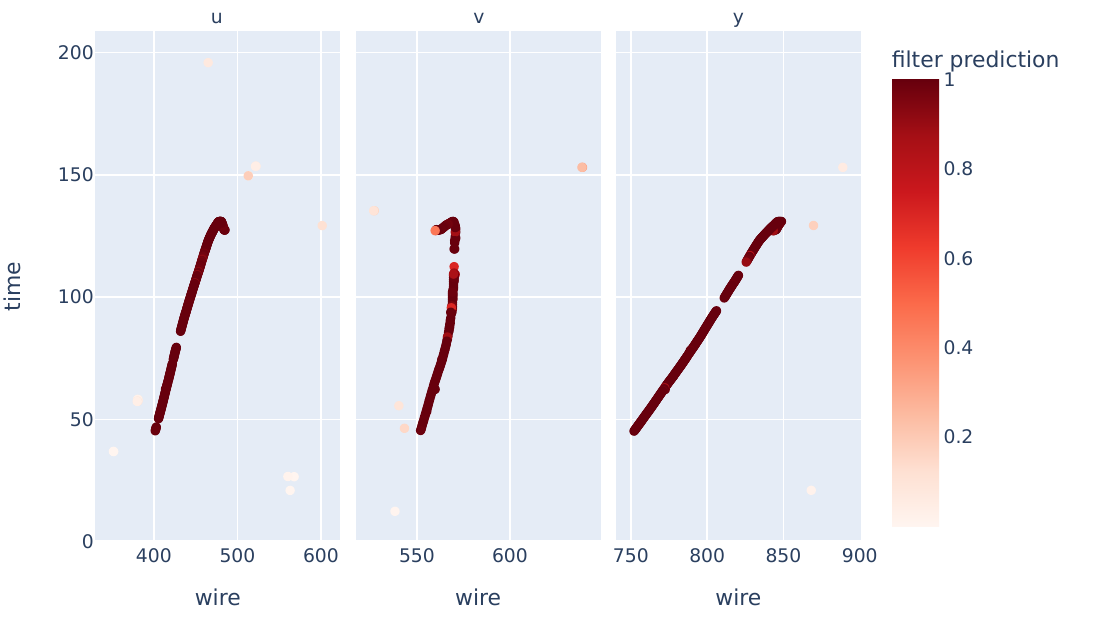}
        \caption{Filter prediction}
    \end{subfigure}
    \begin{subfigure}{0.45\textwidth}
        \includegraphics[width=\textwidth]{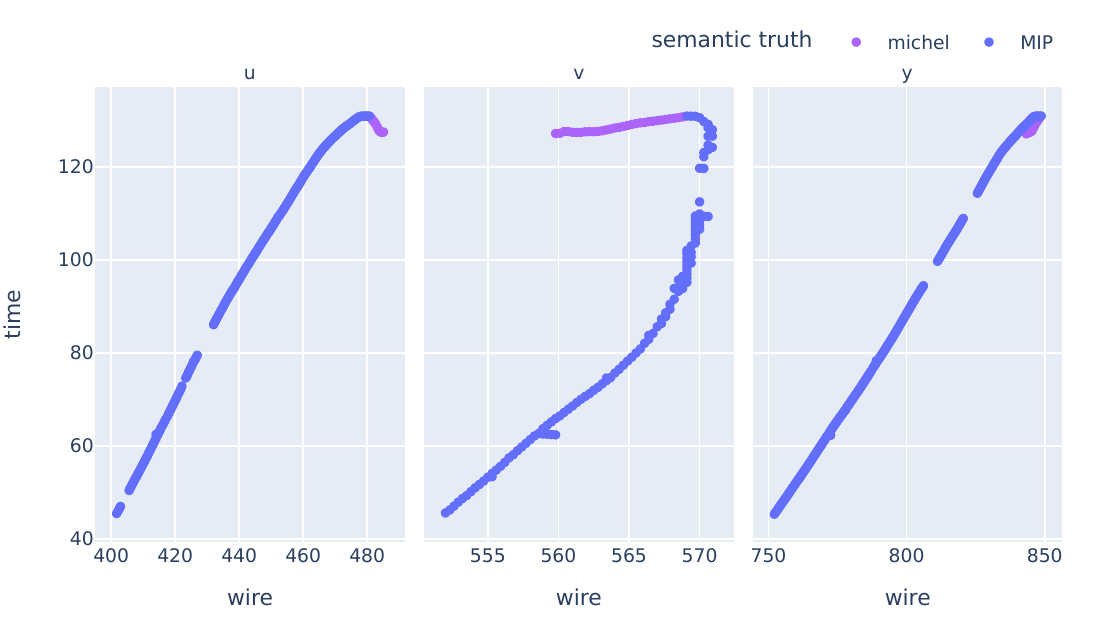}
        \caption{Semantic truth, filtered by truth}
    \end{subfigure}
    \begin{subfigure}{0.45\textwidth}
        \includegraphics[width=\textwidth]{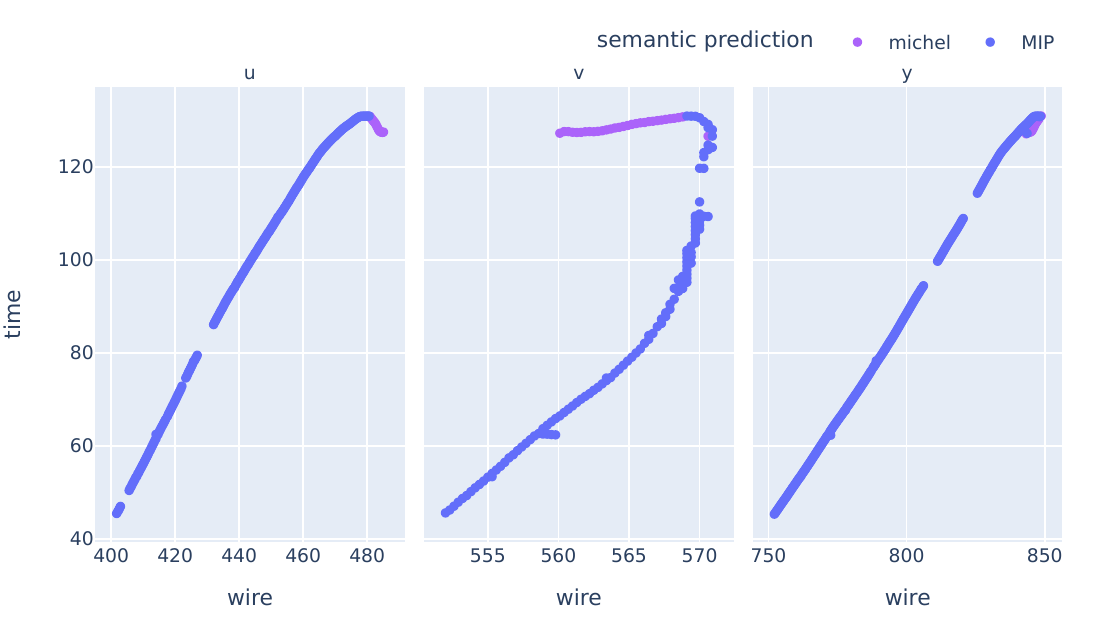}
        \caption{Semantic prediction, filtered by prediction}
    \end{subfigure}
    \caption{Event displays for the third representative event: Run 6110, Subrun 152, Event 7630 of the MicroBooNE open data release.}
    \label{fig:evd-3}
\end{figure}

The third representative event is Run 6110, Subrun 152, Event 7630 of the MicroBooNE open data release, shown in Fig.~\ref{fig:evd-3}. This event is a CCQE $\nu_{\mu}$ interaction consisting of a single track produced by a $\mu^{+}$ with 542~MeV momentum, that decays into a Michel electron with 29~MeV momentum. This event was selected as an example of a well-reconstructed $\nu_{\mu}$ interaction containing a Michel electron.

This event consists of 739 hits across all wire planes, of which 711 are hits from the simulated neutrino interaction. The filter decoder correctly classifies 738 hits, and misclassifies 1 signal hit as background. The semantic decoder misclassifies 3 of the 711 signal hits -- 2 MIP hits are misclassified as Michel, and a single Michel hit is misclassified as MIP.

This interaction is a clean Michel electron topology, and the network is able to reject background hits and classify signal hits with $>99\%$ efficiency.

\subsubsection{Intersecting cosmic}

\begin{figure}
    \centering
    \begin{subfigure}{0.45\textwidth}
        \includegraphics[width=\textwidth]{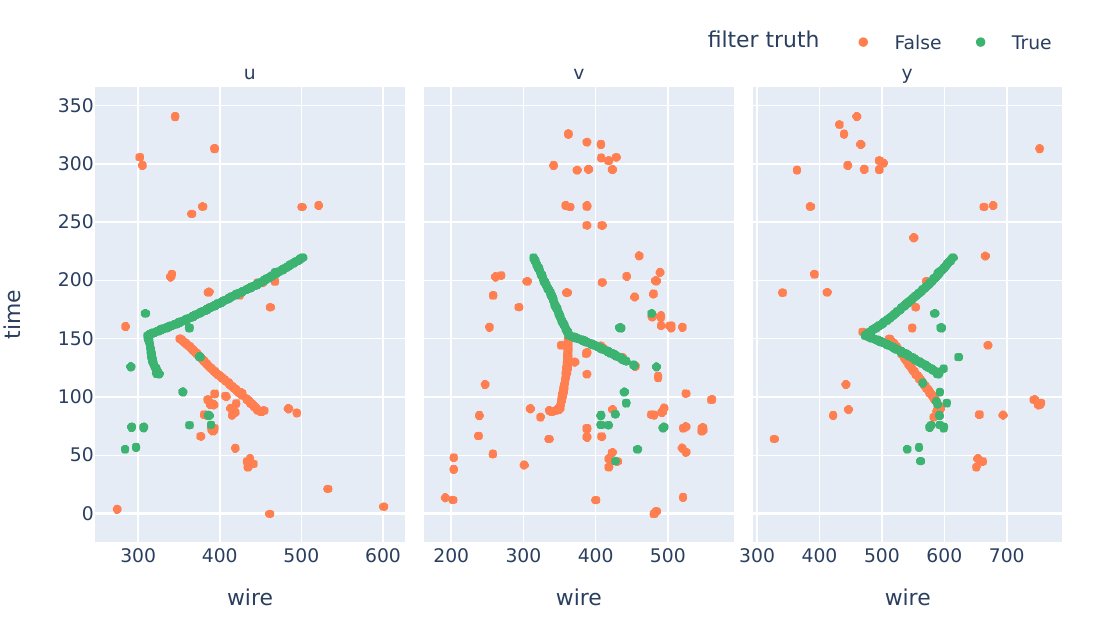}
        \caption{Filter truth}
    \end{subfigure}
    \begin{subfigure}{0.45\textwidth}
        \includegraphics[width=\textwidth]{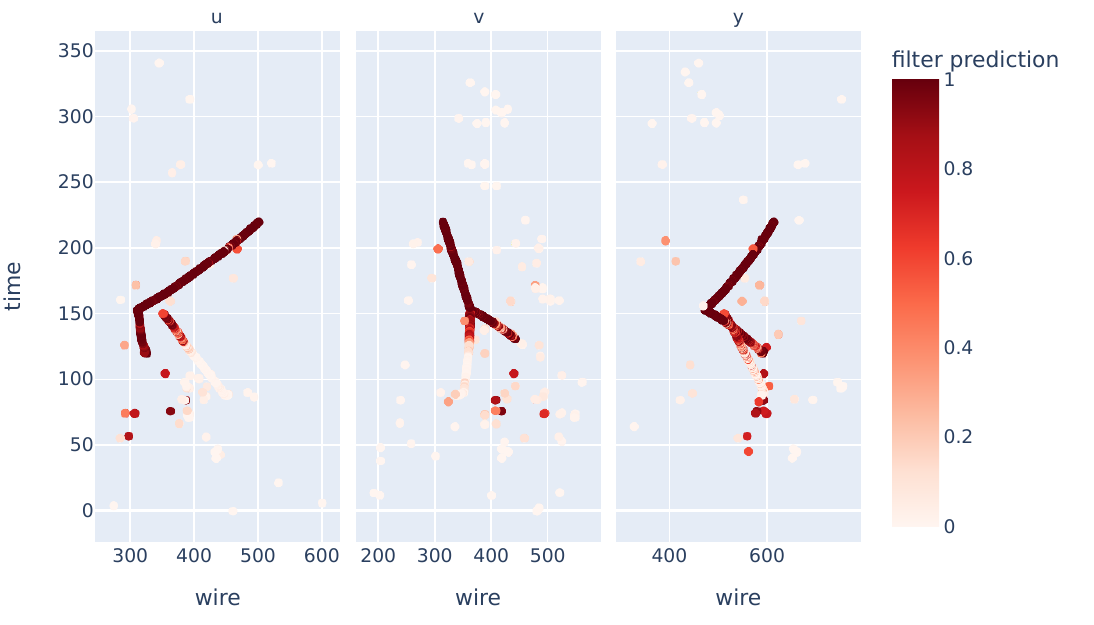}
        \caption{Filter prediction}
    \end{subfigure}
    \begin{subfigure}{0.45\textwidth}
        \includegraphics[width=\textwidth]{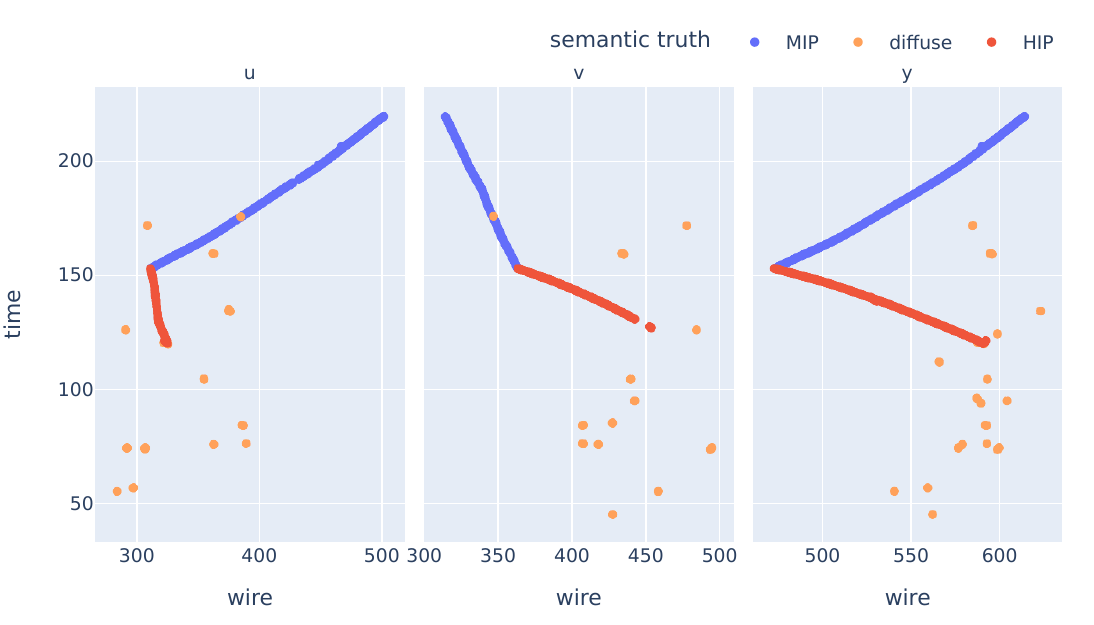}
        \caption{Semantic truth, filtered by truth}
    \end{subfigure}
    \begin{subfigure}{0.45\textwidth}
        \includegraphics[width=\textwidth]{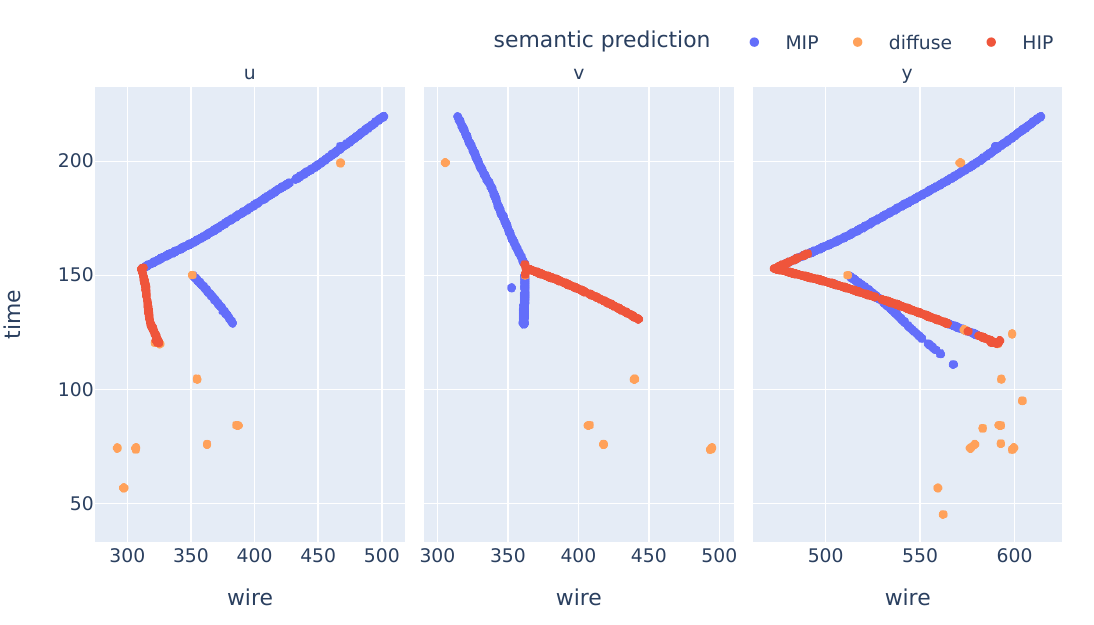}
        \caption{Semantic prediction, filtered by prediction}
    \end{subfigure}
    \caption{Event displays for the fourth representative event: Run 5459, Subrun 94, Event 4738 of the MicroBooNE open data release.}
    \label{fig:evd-4}
\end{figure}

The fourth representative event is Run 5459, Subrun 94, Event 4738 of the MicroBooNE open data release, shown in Fig.~\ref{fig:evd-4}. This event is a CCQE $\nu_{\mu}$ interaction consisting of a longer track produced by a muon with 944~MeV momentum and a shorter track produced by a proton with 1.29~GeV momentum; an unrelated cosmic track also intersects the neutrino event.

This event consists of 3138 hits across all wire planes, of which 2069 are hits from the simulated neutrino interaction. The filter decoder correctly classifies 2771 hits; of the 367 misclassified hits, 118 are signal hits misclassified as background and 249 are background hits misclassified as signal. The semantic decoder misclassifies 133 of the 2069 signal hits -- 36 MIP hits are misclassified as HIP, 83 HIP hits are misclassified (74 as HIP and 9 as diffuse), and 14 diffuse hits are misclassified (13 as MIP and 1 as HIP).

The majority of background hits misclassified as signal by the model belong to the cosmic track that intersects the physics interaction. The proximity of this cosmic track to the simulated neutrino interaction makes it difficult to identify as background, as it points close to the neutrino vertex in 3D. Despite this, the network is able to correctly identify the majority of the track's hits as background, and only misidentifies the track segment closest to the neutrino interaction. The semantic decoder also misclassifies hits at the start of the muon track as being more proton-like, especially in the V and Y planes. It is also worth noting that the network's semantic prediction correctly identifies the background hits from the cosmic track as MIP-like, indicating that the semantic information learned on signal hits from simulation is transferable to background hits from real data.

\subsubsection{High-multiplicity event}

\begin{figure}
    \centering
    \begin{subfigure}{0.45\linewidth}
        \includegraphics[width=\linewidth]{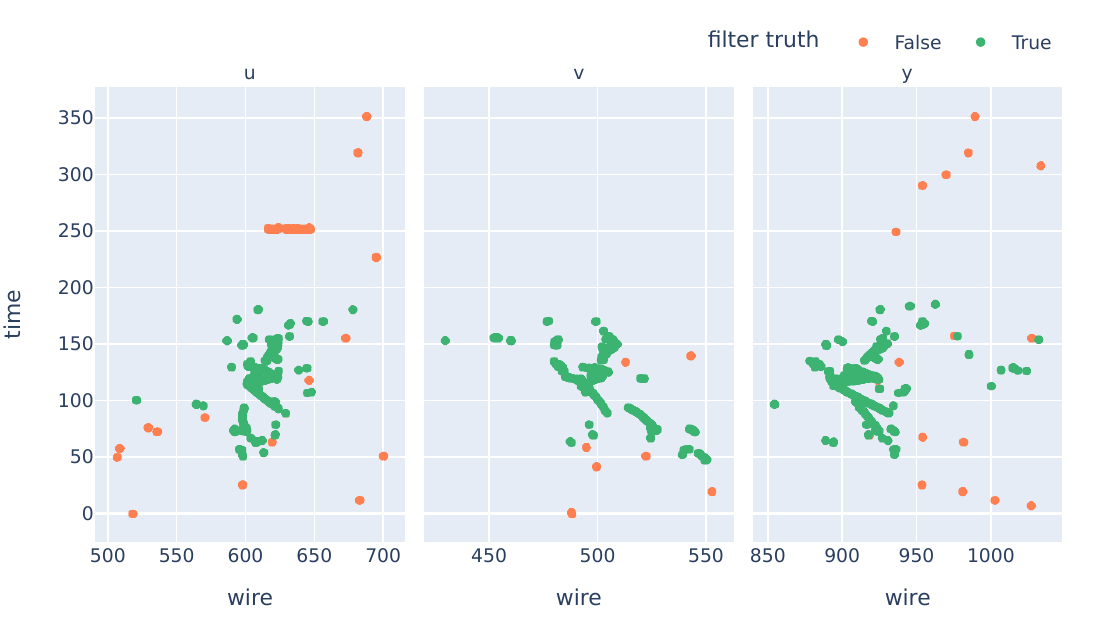}
        \caption{Filter truth}
    \end{subfigure}
    \begin{subfigure}{0.45\linewidth}
        \includegraphics[width=\linewidth]{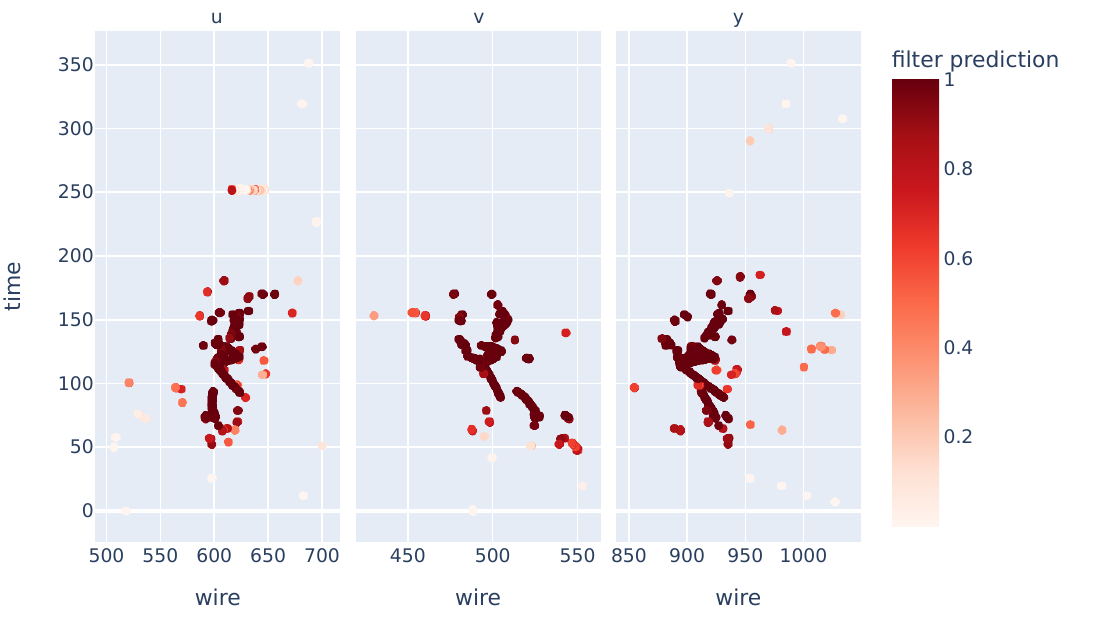}
        \caption{Filter prediction}
    \end{subfigure}
    \begin{subfigure}{0.45\linewidth}
        \includegraphics[width=\linewidth]{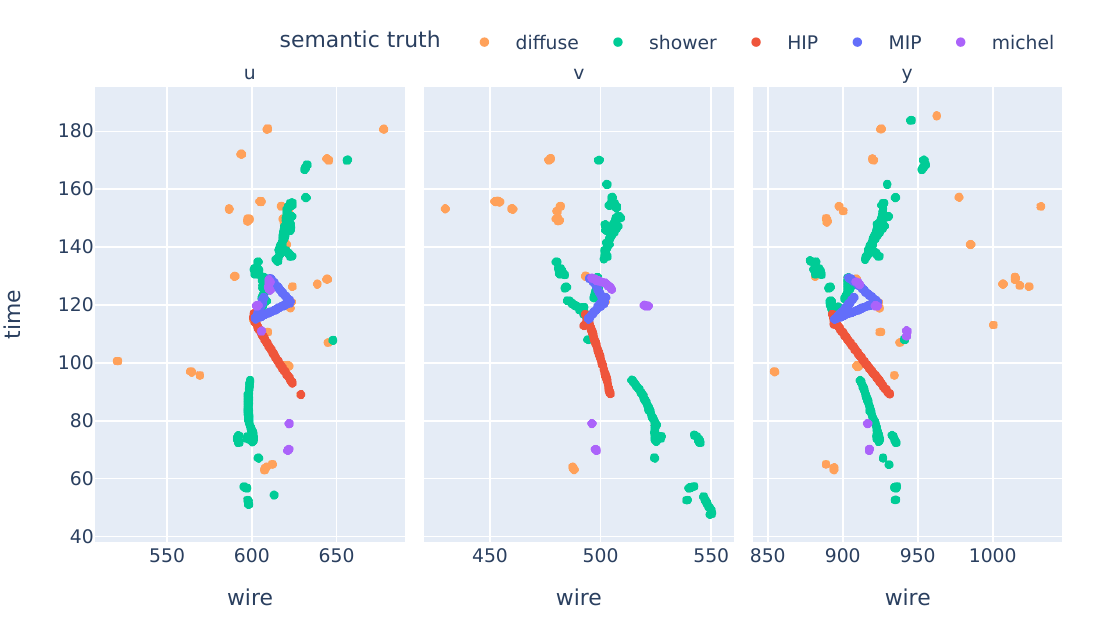}
        \caption{Semantic truth, filtered by truth}
    \end{subfigure}
    \begin{subfigure}{0.45\linewidth}
        \includegraphics[width=\linewidth]{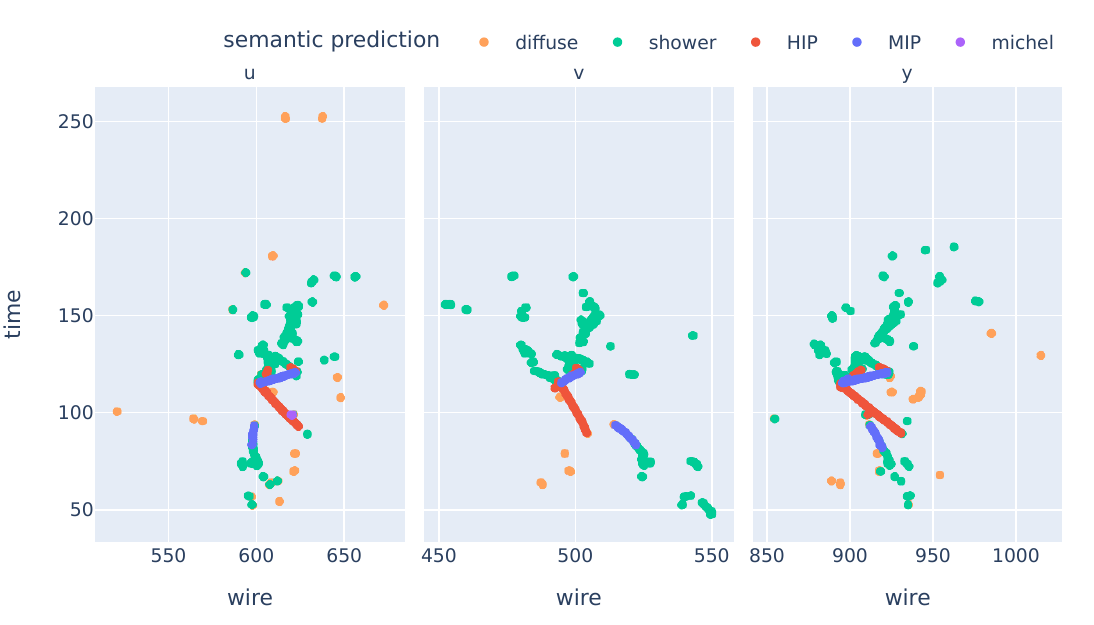}
        \caption{Semantic prediction, filtered by prediction}
    \end{subfigure}
    \caption{Event displays for the fifth representative event: Run 6780, Subrun 200, Event 10006 of the MicroBooNE open data release.}
    \label{fig:evd-5}
\end{figure}

The fifth representative event is Run 6780, Subrun 200, Event 10006 of the MicroBooNE open data release, shown in Fig.~\ref{fig:evd-5}. This event is a CC $\nu_{\mu}$ interaction consisting of a $\mu^{-}$ with momentum~134~MeV, 5 protons with momenta 777~MeV, 145~MeV, 99~MeV, 318~MeV and 177~MeV, two $\pi^{+}$ with momenta 434~MeV and 191~MeV, and two $\pi^{0}$ with momenta 284~MeV and 113~MeV, which are not directly observed but decay into photon pairs with momenta 84~MeV and 231~MeV, and 32~MeV and 144~MeV, respectively. This event was selected as an example of a high-multiplicity event containing hits from all five semantic classes, that is more challenging for the network to reconstruct.

This event consists of 1807 hits across all wire planes, of which 1657 are hits from the simulated neutrino interaction. The filter decoder correctly classifies 1765 hits; of the 42 misclassified hits, 24 are signal hits misclassified as background, and 18 are background hits misclassified as signal. The semantic decoder misclassifies 477 of the 1657 signal hits -- 196 MIP hits (60 as HIP and 136 as shower), 15 HIP hits (14 as shower and 1 as diffuse), 73 shower hits (56 as MIP, 1 as HIP and 16 as diffuse), 88 Michel hits (1 as HIP, 67 as shower and 20 as diffuse) and 105 diffuse hits (2 as HIP, 99 as shower and 4 as Michel).

Events with a higher particle multiplicity are more challenging to reconstruct, as detector hits produced by different particles are more likely to overlap. Additionally, the MicroBooNE open dataset primarily consists of neutrino interactions with a low particle multiplicity, and events with $\geq 10$ primary particles constitute only 10\% of the training set. Considering these limitations, the network performs respectably in correctly semantically labelling over two thirds of the hits in this event.

The 777~MeV momentum proton leaves a long track, which is correctly classified by the model, as is the shorter track produced by a secondary proton with 341~MeV momentum. The primary proton with 318~MeV produces a short track (only 13 hits over all wire planes) and is misclassified by the model as shower-like.

The network performs most poorly on the MIP class. The primary muon track is entirely misclassified, with its hits being assigned to a combination of the shower and HIP classes. The $\pi^{+}$ with 434~MeV momentum is correctly identified as MIP-like; however, it then inelastically scatters to form a new $\pi^{+}$, which decays into a $\mu^{+}$ and finally a Michel positron. These secondary tracks are all misclassified by the model as shower-like, as they overlap with the EM showers produced by the decay of $\pi^{0}$ primaries.

The model identifies the majority of hits belonging to all EM showers, with the exception of one shower root, which it classifies as MIP-like. This is technically correct from a physics perspective, as the showering photon pair-produces into an electron-positron pair that initially appear track-like before undergoing bremsstrahlung and initiating an electromagnetic cascade, but it is not consistent with the semantic truth labelling scheme as currently defined. Such failure modes can likely be resolved in future by redesigning the ground truth labelling scheme to account for them.

\subsubsection{High-multiplicity event}

\begin{figure}
    \centering
    \begin{subfigure}{0.45\linewidth}
        \includegraphics[width=\linewidth]{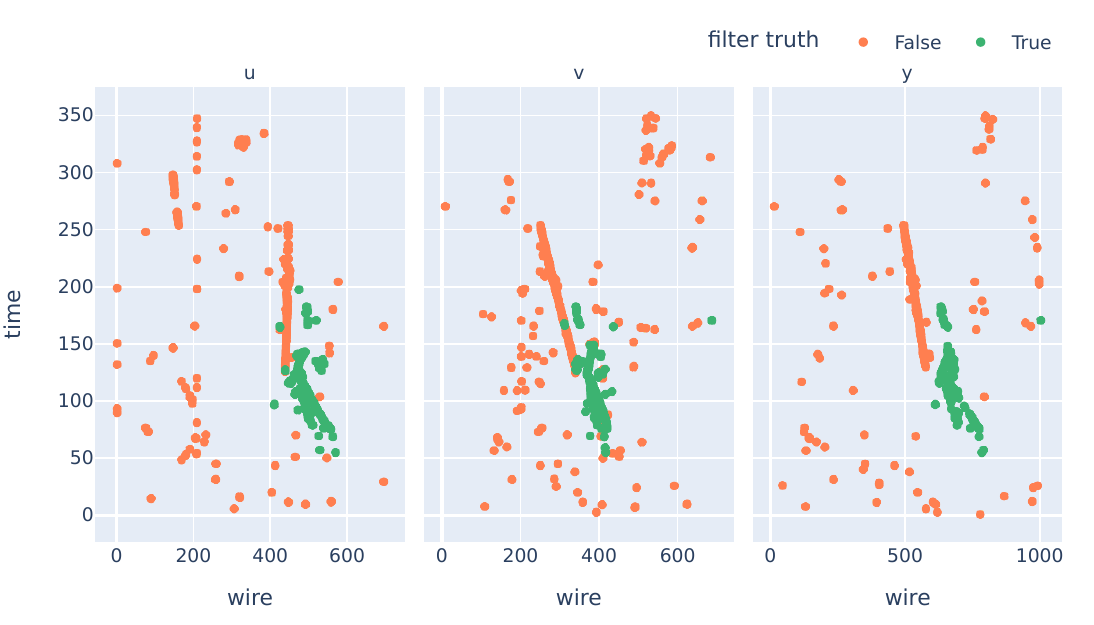}
        \caption{Filter truth}
    \end{subfigure}
    \begin{subfigure}{0.45\linewidth}
        \includegraphics[width=\linewidth]{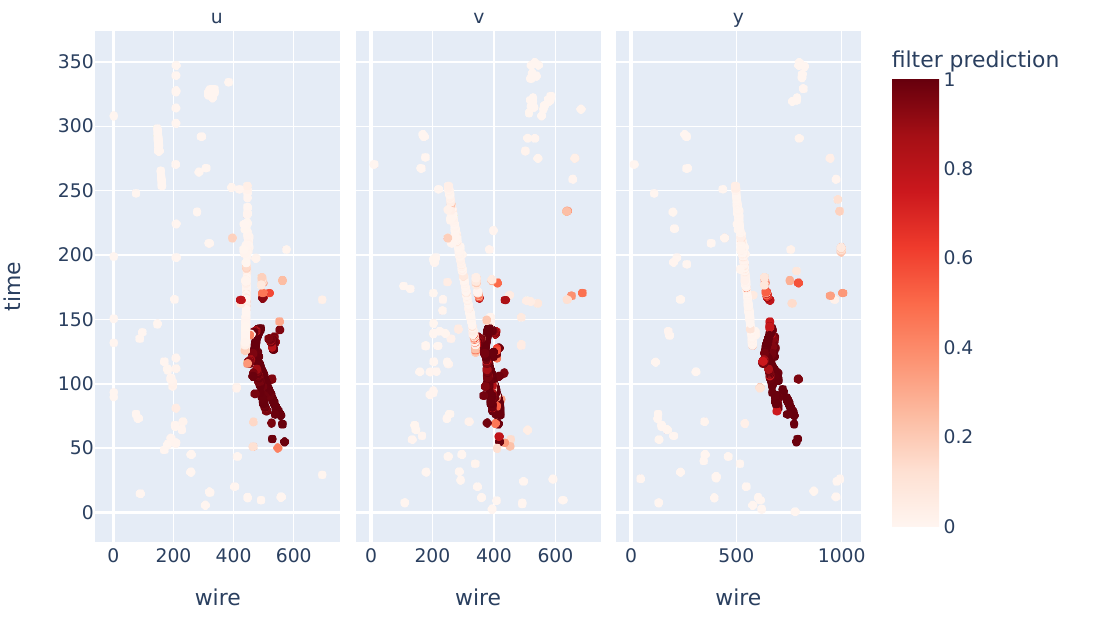}
        \caption{Filter prediction}
    \end{subfigure}
    \begin{subfigure}{0.45\linewidth}
        \includegraphics[width=\linewidth]{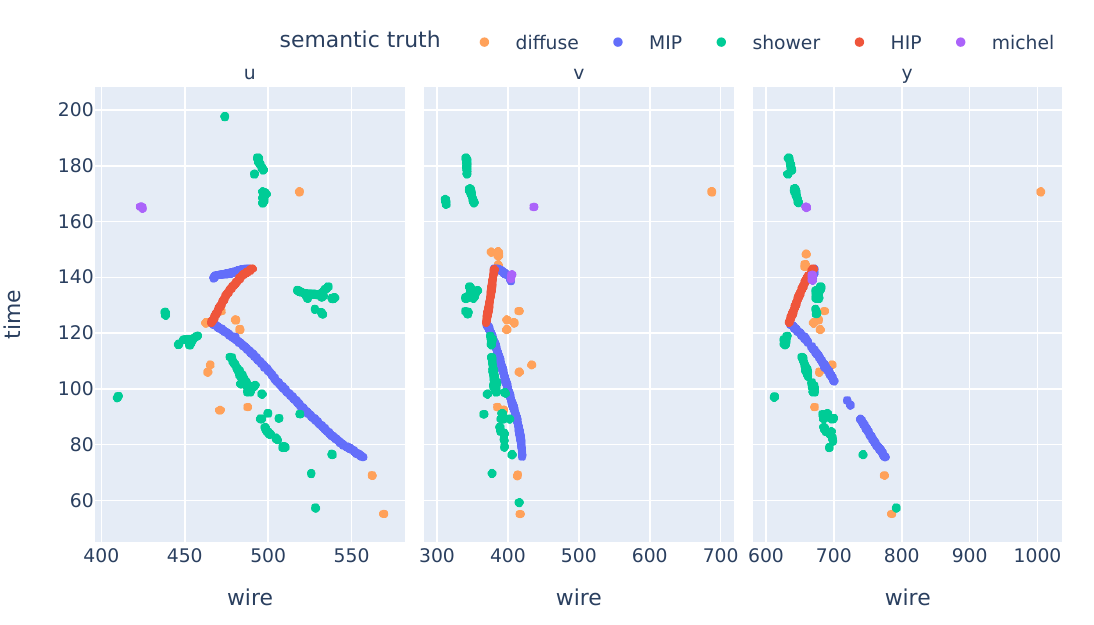}
        \caption{Semantic truth, filtered by truth}
    \end{subfigure}
    \begin{subfigure}{0.45\linewidth}
        \includegraphics[width=\linewidth]{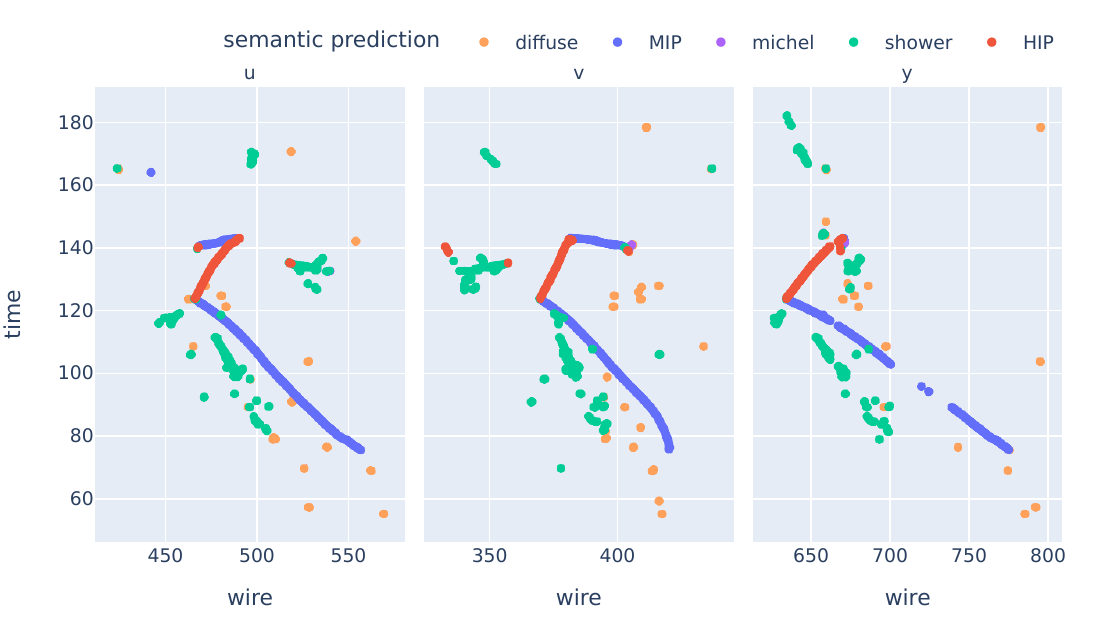}
        \caption{Semantic prediction, filtered by prediction}
    \end{subfigure}
    \caption{Event displays for the sixth representative event: Run 6343, Subrun 72, Event 3606 of the MicroBooNE open data release.}
    \label{fig:evd-6}
\end{figure}

The sixth representative event is Run 6343, Subrun 72, Event 3606 of the MicroBooNE open data release, shown in Fig.~\ref{fig:evd-6}. This event is a CC $\nu_{\mu}$ interaction that produces a $\mu^{-}$ track with 456~MeV momentum, a $\Lambda$ baryon with 701~MeV momentum and a $K^{+}$ meson with 497~MeV momentum. The $\Lambda$ produces neutron activity, a proton track with 169~MeV momentum, and a $\pi^{0}$ with 180~MeV momentum that decays into a photon pair with momenta 178~MeV and 47~MeV. The $K^{+}$ produces a $\pi^{0}$ with 205~MeV momentum that decays into a photon pair with momenta 156~MeV and 90~MeV, and also a $\pi^{+}$ with 205~MeV momentum that decays into a $\mu^{+}$ and then a Michel $e^{+}$. This event was selected as a challenging high-multiplicity interaction on which NuGraph2 demonstrates good performance.

This event consists of 3682 hits across all wire planes, of which 2036 are hits from the simulated neutrino interaction. The filter decoder correctly identifies 3527 hits; of the 155 misclassified hits, 124 are signal hits misclassified as background, and 31 are background hits misclassified as signal. The semantic decoder misclassifies 107 of the 2036 signal hits -- 18 MIP hits are misclassified (5 as HIP, 10 as shower, 1 as Michel and 2 as diffuse), 3 HIP hits are misclassified as shower, 53 shower hits are misclassified (3 as MIP, 8 as HIP, 6 as Michel and 36 as diffuse), 15 Michel hits are misclassified (1 MIP, 1 as HIP, 3 as shower and 10 as diffuse) and 18 diffuse hits are misclassified (17 as shower, 1 as Michel).

The network is able to identify all visible particles in the event, although not all of them are perfectly reconstructed. In regions where energy deposition from EM showers are more sparse, the network exhibits confusion between the shower and diffuse categories. The model correctly identifies the hits associated with the $K^{+}$ track, and the majority of the subsequent $\pi^{+}$ track, but misclassifies some hits in the rapid decay to $\mu^{+}$ and then $e^{+}$, with some hits being identified as shower-like or diffuse. As with the previous example event, these failure modes can potentially be significantly improved upon by
expanding the network to explicitly generate particle clustering as a learning target.

\section{Summary} \label{sec:summary}

NuGraph2 is a graph neural network architecture for particle reconstruction in neutrino physics, achieving 98.0\% efficiency in filtering background hits and 94.9\% efficiency in semantically labelling hits, while producing 2D representations that are 94.8\% consistent. Unlike CNN-based techniques, it is capable of operating on the detector observables in their native structure, without requiring any arbitrary pixelization, voxelization, truncation or downsizing.

The network is also designed to be maximally agnostic to detector technology. It is able to operate on any data structure consisting of detector hits, and offers flexibility in terms of the number of detector planes, truth labelling schemes, and number of graph node input features, rendering it simple to apply to other detector geometries, both LArTPC and otherwise.

NuGraph2's core message-passing engine also provides a general-purpose solution for a variety of tasks. The two decoders presented in this paper are implemented in a modular manner, meaning the architecture can easily be extended or adapted to leverage the core engine for additional learning tasks. Several additional applications are already in active development, utilizing hierarchical graph convolution methods to reconstruct particle and event level quantities, and this work will be discussed in future publications.

\begin{acknowledgments}

This research was supported in part by the U.S. Department of Energy’s Office of Science, Office of High Energy Physics, under Contract No. DE-AC02-07CH11359 (CompHEP Exa.TrkX). 
We would also like to acknowledge the help and support of students working as part of The University of Chicago's Data Science Institute Clinic and Summer Lab programs.
We acknowledge the MicroBooNE Collaboration for making publicly available the data sets~\cite{abratenko_2023_8370883, abratenko_2022_7261921} employed in this work. These data sets consist of simulated neutrino interactions from the Booster Neutrino Beamline overlaid on top of cosmic data collected with the MicroBooNE detector~\cite{MicroBooNE:2016pwy}.

\end{acknowledgments}

\appendix

\bibliography{nugraph2}

\providecommand{\noopsort}[1]{}\providecommand{\singleletter}[1]{#1}%
\begin{thebibliography}{37}%
\makeatletter
\providecommand \@ifxundefined [1]{%
 \@ifx{#1\undefined}
}%
\providecommand \@ifnum [1]{%
 \ifnum #1\expandafter \@firstoftwo
 \else \expandafter \@secondoftwo
 \fi
}%
\providecommand \@ifx [1]{%
 \ifx #1\expandafter \@firstoftwo
 \else \expandafter \@secondoftwo
 \fi
}%
\providecommand \natexlab [1]{#1}%
\providecommand \enquote  [1]{``#1''}%
\providecommand \bibnamefont  [1]{#1}%
\providecommand \bibfnamefont [1]{#1}%
\providecommand \citenamefont [1]{#1}%
\providecommand \href@noop [0]{\@secondoftwo}%
\providecommand \href [0]{\begingroup \@sanitize@url \@href}%
\providecommand \@href[1]{\@@startlink{#1}\@@href}%
\providecommand \@@href[1]{\endgroup#1\@@endlink}%
\providecommand \@sanitize@url [0]{\catcode `\\12\catcode `\$12\catcode
  `\&12\catcode `\#12\catcode `\^12\catcode `\_12\catcode `\%12\relax}%
\providecommand \@@startlink[1]{}%
\providecommand \@@endlink[0]{}%
\providecommand \url  [0]{\begingroup\@sanitize@url \@url }%
\providecommand \@url [1]{\endgroup\@href {#1}{\urlprefix }}%
\providecommand \urlprefix  [0]{URL }%
\providecommand \Eprint [0]{\href }%
\providecommand \doibase [0]{https://doi.org/}%
\providecommand \selectlanguage [0]{\@gobble}%
\providecommand \bibinfo  [0]{\@secondoftwo}%
\providecommand \bibfield  [0]{\@secondoftwo}%
\providecommand \translation [1]{[#1]}%
\providecommand \BibitemOpen [0]{}%
\providecommand \bibitemStop [0]{}%
\providecommand \bibitemNoStop [0]{.\EOS\space}%
\providecommand \EOS [0]{\spacefactor3000\relax}%
\providecommand \BibitemShut  [1]{\csname bibitem#1\endcsname}%
\let\auto@bib@innerbib\@empty
\bibitem [{\citenamefont {Aurisano}\ \emph {et~al.}(2016)\citenamefont
  {Aurisano}, \citenamefont {Radovic}, \citenamefont {Rocco}, \citenamefont
  {Himmel}, \citenamefont {Messier}, \citenamefont {Niner}, \citenamefont
  {Pawloski}, \citenamefont {Psihas}, \citenamefont {Sousa},\ and\
  \citenamefont {Vahle}}]{Aurisano_2016}%
  \BibitemOpen
  \bibfield  {author} {\bibinfo {author} {\bibfnamefont {A.}~\bibnamefont
  {Aurisano}}, \bibinfo {author} {\bibfnamefont {A.}~\bibnamefont {Radovic}},
  \bibinfo {author} {\bibfnamefont {D.}~\bibnamefont {Rocco}}, \bibinfo
  {author} {\bibfnamefont {A.}~\bibnamefont {Himmel}}, \bibinfo {author}
  {\bibfnamefont {M.}~\bibnamefont {Messier}}, \bibinfo {author} {\bibfnamefont
  {E.}~\bibnamefont {Niner}}, \bibinfo {author} {\bibfnamefont
  {G.}~\bibnamefont {Pawloski}}, \bibinfo {author} {\bibfnamefont
  {F.}~\bibnamefont {Psihas}}, \bibinfo {author} {\bibfnamefont
  {A.}~\bibnamefont {Sousa}},\ and\ \bibinfo {author} {\bibfnamefont
  {P.}~\bibnamefont {Vahle}},\ }\bibfield  {title} {\bibinfo {title} {A
  convolutional neural network neutrino event classifier},\ }\href
  {https://doi.org/10.1088/1748-0221/11/09/p09001} {\bibfield  {journal}
  {\bibinfo  {journal} {Journal of Instrumentation}\ }\textbf {\bibinfo
  {volume} {11}}\bibinfo  {number} { (09)},\ \bibinfo {pages}
  {P09001}}\BibitemShut {NoStop}%
\bibitem [{\citenamefont {Psihas}\ \emph {et~al.}(2019)\citenamefont {Psihas},
  \citenamefont {Niner}, \citenamefont {Groh}, \citenamefont {Murphy},
  \citenamefont {Aurisano}, \citenamefont {Himmel}, \citenamefont {Lang},
  \citenamefont {Messier}, \citenamefont {Radovic},\ and\ \citenamefont
  {Sousa}}]{Psihas:2019ksa}%
  \BibitemOpen
\bibfield  {number} {  }\bibfield  {author} {\bibinfo {author} {\bibfnamefont
  {F.}~\bibnamefont {Psihas}}, \bibinfo {author} {\bibfnamefont
  {E.}~\bibnamefont {Niner}}, \bibinfo {author} {\bibfnamefont
  {M.}~\bibnamefont {Groh}}, \bibinfo {author} {\bibfnamefont {R.}~\bibnamefont
  {Murphy}}, \bibinfo {author} {\bibfnamefont {A.}~\bibnamefont {Aurisano}},
  \bibinfo {author} {\bibfnamefont {A.}~\bibnamefont {Himmel}}, \bibinfo
  {author} {\bibfnamefont {K.}~\bibnamefont {Lang}}, \bibinfo {author}
  {\bibfnamefont {M.~D.}\ \bibnamefont {Messier}}, \bibinfo {author}
  {\bibfnamefont {A.}~\bibnamefont {Radovic}},\ and\ \bibinfo {author}
  {\bibfnamefont {A.}~\bibnamefont {Sousa}},\ }\bibfield  {title} {\bibinfo
  {title} {{Context-Enriched Identification of Particles with a Convolutional
  Network for Neutrino Events}},\ }\href
  {https://doi.org/10.1103/PhysRevD.100.073005} {\bibfield  {journal} {\bibinfo
   {journal} {Phys. Rev. D}\ }\textbf {\bibinfo {volume} {100}},\ \bibinfo
  {pages} {073005} (\bibinfo {year} {2019})},\ \Eprint
  {https://arxiv.org/abs/1906.00713} {arXiv:1906.00713 [physics.ins-det]}
  \BibitemShut {NoStop}%
\bibitem [{\citenamefont {Abi}\ \emph {et~al.}(2020{\natexlab{a}})\citenamefont
  {Abi} \emph {et~al.}}]{DUNE:2020gpm}%
  \BibitemOpen
  \bibfield  {author} {\bibinfo {author} {\bibfnamefont {B.}~\bibnamefont
  {Abi}} \emph {et~al.} (\bibinfo {collaboration} {DUNE}),\ }\bibfield  {title}
  {\bibinfo {title} {{Neutrino interaction classification with a convolutional
  neural network in the DUNE far detector}},\ }\href
  {https://doi.org/10.1103/PhysRevD.102.092003} {\bibfield  {journal} {\bibinfo
   {journal} {Phys. Rev. D}\ }\textbf {\bibinfo {volume} {102}},\ \bibinfo
  {pages} {092003} (\bibinfo {year} {2020}{\natexlab{a}})},\ \Eprint
  {https://arxiv.org/abs/2006.15052} {arXiv:2006.15052 [physics.ins-det]}
  \BibitemShut {NoStop}%
\bibitem [{\citenamefont {Abed~Abud}\ \emph {et~al.}(2022)\citenamefont
  {Abed~Abud} \emph {et~al.}}]{DUNE:2022fiy}%
  \BibitemOpen
  \bibfield  {author} {\bibinfo {author} {\bibfnamefont {A.}~\bibnamefont
  {Abed~Abud}} \emph {et~al.} (\bibinfo {collaboration} {DUNE}),\ }\bibfield
  {title} {\bibinfo {title} {{Separation of track- and shower-like energy
  deposits in ProtoDUNE-SP using a convolutional neural network}},\ }\href
  {https://doi.org/10.1140/epjc/s10052-022-10791-2} {\bibfield  {journal}
  {\bibinfo  {journal} {Eur. Phys. J. C}\ }\textbf {\bibinfo {volume} {82}},\
  \bibinfo {pages} {903} (\bibinfo {year} {2022})},\ \Eprint
  {https://arxiv.org/abs/2203.17053} {arXiv:2203.17053 [physics.ins-det]}
  \BibitemShut {NoStop}%
\bibitem [{\citenamefont {Adams}\ \emph {et~al.}(2019)\citenamefont {Adams}
  \emph {et~al.}}]{MicroBooNE:2018kka}%
  \BibitemOpen
  \bibfield  {author} {\bibinfo {author} {\bibfnamefont {C.}~\bibnamefont
  {Adams}} \emph {et~al.} (\bibinfo {collaboration} {MicroBooNE}),\ }\bibfield
  {title} {\bibinfo {title} {{Deep neural network for pixel-level
  electromagnetic particle identification in the MicroBooNE liquid argon time
  projection chamber}},\ }\href {https://doi.org/10.1103/PhysRevD.99.092001}
  {\bibfield  {journal} {\bibinfo  {journal} {Phys. Rev. D}\ }\textbf {\bibinfo
  {volume} {99}},\ \bibinfo {pages} {092001} (\bibinfo {year} {2019})},\
  \Eprint {https://arxiv.org/abs/1808.07269} {arXiv:1808.07269 [hep-ex]}
  \BibitemShut {NoStop}%
\bibitem [{\citenamefont {Abratenko}\ \emph
  {et~al.}(2021{\natexlab{a}})\citenamefont {Abratenko} \emph
  {et~al.}}]{MicroBooNE:2020hho}%
  \BibitemOpen
  \bibfield  {author} {\bibinfo {author} {\bibfnamefont {P.}~\bibnamefont
  {Abratenko}} \emph {et~al.} (\bibinfo {collaboration} {MicroBooNE}),\
  }\bibfield  {title} {\bibinfo {title} {{Convolutional neural network for
  multiple particle identification in the MicroBooNE liquid argon time
  projection chamber}},\ }\href {https://doi.org/10.1103/PhysRevD.103.092003}
  {\bibfield  {journal} {\bibinfo  {journal} {Phys. Rev. D}\ }\textbf {\bibinfo
  {volume} {103}},\ \bibinfo {pages} {092003} (\bibinfo {year}
  {2021}{\natexlab{a}})},\ \Eprint {https://arxiv.org/abs/2010.08653}
  {arXiv:2010.08653 [hep-ex]} \BibitemShut {NoStop}%
\bibitem [{\citenamefont {Choy}\ \emph {et~al.}(2019)\citenamefont {Choy},
  \citenamefont {Gwak},\ and\ \citenamefont {Savarese}}]{choy20194d}%
  \BibitemOpen
  \bibfield  {author} {\bibinfo {author} {\bibfnamefont {C.}~\bibnamefont
  {Choy}}, \bibinfo {author} {\bibfnamefont {J.}~\bibnamefont {Gwak}},\ and\
  \bibinfo {author} {\bibfnamefont {S.}~\bibnamefont {Savarese}},\ }\bibfield
  {title} {\bibinfo {title} {4d spatio-temporal convnets: Minkowski
  convolutional neural networks},\ }in\ \href@noop {} {\emph {\bibinfo
  {booktitle} {Proceedings of the IEEE Conference on Computer Vision and
  Pattern Recognition}}}\ (\bibinfo {year} {2019})\ pp.\ \bibinfo {pages}
  {3075--3084}\BibitemShut {NoStop}%
\bibitem [{\citenamefont {Koh}\ \emph {et~al.}(2020)\citenamefont {Koh},
  \citenamefont {de~Soux}, \citenamefont {Dominé}, \citenamefont {Drielsma},
  \citenamefont {Itay}, \citenamefont {Lin}, \citenamefont {Terao},
  \citenamefont {Tsang},\ and\ \citenamefont {Usher}}]{Koh:2020snv}%
  \BibitemOpen
  \bibfield  {author} {\bibinfo {author} {\bibfnamefont {D.~H.}\ \bibnamefont
  {Koh}}, \bibinfo {author} {\bibfnamefont {P.~C.}\ \bibnamefont {de~Soux}},
  \bibinfo {author} {\bibfnamefont {L.}~\bibnamefont {Dominé}}, \bibinfo
  {author} {\bibfnamefont {F.}~\bibnamefont {Drielsma}}, \bibinfo {author}
  {\bibfnamefont {R.}~\bibnamefont {Itay}}, \bibinfo {author} {\bibfnamefont
  {Q.}~\bibnamefont {Lin}}, \bibinfo {author} {\bibfnamefont {K.}~\bibnamefont
  {Terao}}, \bibinfo {author} {\bibfnamefont {K.~V.}\ \bibnamefont {Tsang}},\
  and\ \bibinfo {author} {\bibfnamefont {T.}~\bibnamefont {Usher}},\
  }\href@noop {} {\bibinfo {title} {Scalable, proposal-free instance
  segmentation network for 3d pixel clustering and particle trajectory
  reconstruction in liquid argon time projection chambers}} (\bibinfo {year}
  {2020}),\ \Eprint {https://arxiv.org/abs/2007.03083} {arXiv:2007.03083
  [physics.ins-det]} \BibitemShut {NoStop}%
\bibitem [{\citenamefont {Abratenko}\ \emph
  {et~al.}(2021{\natexlab{b}})\citenamefont {Abratenko} \emph
  {et~al.}}]{MicroBooNE:2020yze}%
  \BibitemOpen
  \bibfield  {author} {\bibinfo {author} {\bibfnamefont {P.}~\bibnamefont
  {Abratenko}} \emph {et~al.} (\bibinfo {collaboration} {MicroBooNE}),\
  }\bibfield  {title} {\bibinfo {title} {{Semantic segmentation with a sparse
  convolutional neural network for event reconstruction in MicroBooNE}},\
  }\href {https://doi.org/10.1103/PhysRevD.103.052012} {\bibfield  {journal}
  {\bibinfo  {journal} {Phys. Rev. D}\ }\textbf {\bibinfo {volume} {103}},\
  \bibinfo {pages} {052012} (\bibinfo {year} {2021}{\natexlab{b}})},\ \Eprint
  {https://arxiv.org/abs/2012.08513} {arXiv:2012.08513 [physics.ins-det]}
  \BibitemShut {NoStop}%
\bibitem [{\citenamefont {Abed~Abud}(2023)}]{AbedAbud:2023cmz}%
  \BibitemOpen
  \bibfield  {author} {\bibinfo {author} {\bibfnamefont {A.}~\bibnamefont
  {Abed~Abud}} (\bibinfo {collaboration} {DUNE}),\ }\bibfield  {title}
  {\bibinfo {title} {{Sparse Convolutional Neural Networks for particle
  classification in ProtoDUNE-SP events}},\ }\href
  {https://doi.org/10.1088/1742-6596/2438/1/012125} {\bibfield  {journal}
  {\bibinfo  {journal} {J. Phys. Conf. Ser.}\ }\textbf {\bibinfo {volume}
  {2438}},\ \bibinfo {pages} {012125} (\bibinfo {year} {2023})}\BibitemShut
  {NoStop}%
\bibitem [{\citenamefont {Farrell}\ \emph {et~al.}(2018)\citenamefont {Farrell}
  \emph {et~al.}}]{Farrell:2018cjr}%
  \BibitemOpen
  \bibfield  {author} {\bibinfo {author} {\bibfnamefont {S.}~\bibnamefont
  {Farrell}} \emph {et~al.},\ }\bibfield  {title} {\bibinfo {title} {{Novel
  deep learning methods for track reconstruction}},\ }in\ \href@noop {} {\emph
  {\bibinfo {booktitle} {{4th International Workshop Connecting The Dots
  2018}}}}\ (\bibinfo {year} {2018})\ \Eprint
  {https://arxiv.org/abs/1810.06111} {arXiv:1810.06111 [hep-ex]} \BibitemShut
  {NoStop}%
\bibitem [{\citenamefont {Ju}\ \emph {et~al.}(2021)\citenamefont {Ju} \emph
  {et~al.}}]{ExaTrkX:2021abe}%
  \BibitemOpen
  \bibfield  {author} {\bibinfo {author} {\bibfnamefont {X.}~\bibnamefont {Ju}}
  \emph {et~al.} (\bibinfo {collaboration} {Exa.TrkX}),\ }\bibfield  {title}
  {\bibinfo {title} {{Performance of a geometric deep learning pipeline for
  HL-LHC particle tracking}},\ }\href
  {https://doi.org/10.1140/epjc/s10052-021-09675-8} {\bibfield  {journal}
  {\bibinfo  {journal} {Eur. Phys. J. C}\ }\textbf {\bibinfo {volume} {81}},\
  \bibinfo {pages} {876} (\bibinfo {year} {2021})},\ \Eprint
  {https://arxiv.org/abs/2103.06995} {arXiv:2103.06995 [physics.data-an]}
  \BibitemShut {NoStop}%
\bibitem [{\citenamefont {Liu}\ \emph {et~al.}(2023)\citenamefont {Liu},
  \citenamefont {Calafiura}, \citenamefont {Farrell}, \citenamefont {Ju},
  \citenamefont {Murnane},\ and\ \citenamefont {Pham}}]{Liu:2023siw}%
  \BibitemOpen
  \bibfield  {author} {\bibinfo {author} {\bibfnamefont {R.}~\bibnamefont
  {Liu}}, \bibinfo {author} {\bibfnamefont {P.}~\bibnamefont {Calafiura}},
  \bibinfo {author} {\bibfnamefont {S.}~\bibnamefont {Farrell}}, \bibinfo
  {author} {\bibfnamefont {X.}~\bibnamefont {Ju}}, \bibinfo {author}
  {\bibfnamefont {D.~T.}\ \bibnamefont {Murnane}},\ and\ \bibinfo {author}
  {\bibfnamefont {T.~M.}\ \bibnamefont {Pham}},\ }\bibfield  {title} {\bibinfo
  {title} {{Hierarchical Graph Neural Networks for Particle Track
  Reconstruction}},\ }in\ \href@noop {} {\emph {\bibinfo {booktitle} {{21th
  International Workshop on Advanced Computing and Analysis Techniques in
  Physics Research}: {AI meets Reality}}}}\ (\bibinfo {year} {2023})\ \Eprint
  {https://arxiv.org/abs/2303.01640} {arXiv:2303.01640 [hep-ex]} \BibitemShut
  {NoStop}%
\bibitem [{\citenamefont {Murnane}\ \emph {et~al.}(2023)\citenamefont
  {Murnane}, \citenamefont {Thais},\ and\ \citenamefont
  {Thete}}]{Murnane:2023kfm}%
  \BibitemOpen
  \bibfield  {author} {\bibinfo {author} {\bibfnamefont {D.}~\bibnamefont
  {Murnane}}, \bibinfo {author} {\bibfnamefont {S.}~\bibnamefont {Thais}},\
  and\ \bibinfo {author} {\bibfnamefont {A.}~\bibnamefont {Thete}},\ }\bibfield
   {title} {\bibinfo {title} {{Equivariant Graph Neural Networks for Charged
  Particle Tracking}},\ }in\ \href@noop {} {\emph {\bibinfo {booktitle} {{21th
  International Workshop on Advanced Computing and Analysis Techniques in
  Physics Research}: {AI meets Reality}}}}\ (\bibinfo {year} {2023})\ \Eprint
  {https://arxiv.org/abs/2304.05293} {arXiv:2304.05293 [physics.ins-det]}
  \BibitemShut {NoStop}%
\bibitem [{\citenamefont {Lieret}\ and\ \citenamefont
  {DeZoort}(2023)}]{Lieret:2023ydc}%
  \BibitemOpen
  \bibfield  {author} {\bibinfo {author} {\bibfnamefont {K.}~\bibnamefont
  {Lieret}}\ and\ \bibinfo {author} {\bibfnamefont {G.}~\bibnamefont
  {DeZoort}},\ }\bibfield  {title} {\bibinfo {title} {{An Object Condensation
  Pipeline for Charged Particle Tracking at the High Luminosity LHC}},\ }in\
  \href@noop {} {\emph {\bibinfo {booktitle} {{26th International Conference on
  Computing in High Energy \& Nuclear Physics}}}}\ (\bibinfo {year} {2023})\
  \Eprint {https://arxiv.org/abs/2309.16754} {arXiv:2309.16754
  [physics.data-an]} \BibitemShut {NoStop}%
\bibitem [{\citenamefont {Choma}\ \emph {et~al.}(2018)\citenamefont {Choma},
  \citenamefont {Monti}, \citenamefont {Gerhardt}, \citenamefont {Palczewski},
  \citenamefont {Ronaghi}, \citenamefont {Prabhat}, \citenamefont {Bhimji},
  \citenamefont {Bronstein}, \citenamefont {Klein},\ and\ \citenamefont
  {Bruna}}]{8614089}%
  \BibitemOpen
  \bibfield  {author} {\bibinfo {author} {\bibfnamefont {N.}~\bibnamefont
  {Choma}}, \bibinfo {author} {\bibfnamefont {F.}~\bibnamefont {Monti}},
  \bibinfo {author} {\bibfnamefont {L.}~\bibnamefont {Gerhardt}}, \bibinfo
  {author} {\bibfnamefont {T.}~\bibnamefont {Palczewski}}, \bibinfo {author}
  {\bibfnamefont {Z.}~\bibnamefont {Ronaghi}}, \bibinfo {author} {\bibfnamefont
  {P.}~\bibnamefont {Prabhat}}, \bibinfo {author} {\bibfnamefont
  {W.}~\bibnamefont {Bhimji}}, \bibinfo {author} {\bibfnamefont {M.~M.}\
  \bibnamefont {Bronstein}}, \bibinfo {author} {\bibfnamefont {S.~R.}\
  \bibnamefont {Klein}},\ and\ \bibinfo {author} {\bibfnamefont
  {J.}~\bibnamefont {Bruna}},\ }\bibfield  {title} {\bibinfo {title} {Graph
  neural networks for icecube signal classification},\ }in\ \href
  {https://doi.org/10.1109/ICMLA.2018.00064} {\emph {\bibinfo {booktitle} {2018
  17th IEEE International Conference on Machine Learning and Applications
  (ICMLA)}}}\ (\bibinfo {year} {2018})\ pp.\ \bibinfo {pages}
  {386--391}\BibitemShut {NoStop}%
\bibitem [{\citenamefont {Drielsma}\ \emph {et~al.}(2021)\citenamefont
  {Drielsma}, \citenamefont {Terao}, \citenamefont {Domin\'e},\ and\
  \citenamefont {Koh}}]{Drielsma:2021jdv}%
  \BibitemOpen
  \bibfield  {author} {\bibinfo {author} {\bibfnamefont {F.}~\bibnamefont
  {Drielsma}}, \bibinfo {author} {\bibfnamefont {K.}~\bibnamefont {Terao}},
  \bibinfo {author} {\bibfnamefont {L.}~\bibnamefont {Domin\'e}},\ and\
  \bibinfo {author} {\bibfnamefont {D.~H.}\ \bibnamefont {Koh}},\ }\bibfield
  {title} {\bibinfo {title} {{Scalable, End-to-End, Deep-Learning-Based Data
  Reconstruction Chain for Particle Imaging Detectors}},\ }in\ \href@noop {}
  {\emph {\bibinfo {booktitle} {{34th Conference on Neural Information
  Processing Systems}}}}\ (\bibinfo {year} {2021})\ \Eprint
  {https://arxiv.org/abs/2102.01033} {arXiv:2102.01033 [hep-ex]} \BibitemShut
  {NoStop}%
\bibitem [{\citenamefont {Hewes}\ \emph {et~al.}(2021)\citenamefont {Hewes}
  \emph {et~al.}}]{Hewes_2021}%
  \BibitemOpen
  \bibfield  {author} {\bibinfo {author} {\bibfnamefont {V.}~\bibnamefont
  {Hewes}} \emph {et~al.} (\bibinfo {collaboration} {Exa.TrkX}),\ }\bibfield
  {title} {\bibinfo {title} {Graph neural network for object reconstruction in
  liquid argon time projection chambers},\ }\href
  {https://doi.org/10.1051/epjconf/202125103054} {\bibfield  {journal}
  {\bibinfo  {journal} {{EPJ} Web of Conferences}\ }\textbf {\bibinfo {volume}
  {251}},\ \bibinfo {pages} {03054} (\bibinfo {year} {2021})}\BibitemShut
  {NoStop}%
\bibitem [{\citenamefont {Cerati}(2023)}]{Cerati:2023rtv}%
  \BibitemOpen
  \bibfield  {author} {\bibinfo {author} {\bibfnamefont {G.}~\bibnamefont
  {Cerati}} (\bibinfo {collaboration} {MicroBooNE}),\ }\bibfield  {title}
  {\bibinfo {title} {{MicroBooNE Public Data Sets: a Collaborative Tool for
  LArTPC Software Development}},\ }in\ \href@noop {} {\emph {\bibinfo
  {booktitle} {{26th International Conference on Computing in High Energy \&
  Nuclear Physics}}}}\ (\bibinfo {year} {2023})\ \Eprint
  {https://arxiv.org/abs/2309.15362} {arXiv:2309.15362 [hep-ex]} \BibitemShut
  {NoStop}%
\bibitem [{\citenamefont {Acciarri}\ \emph {et~al.}(2017)\citenamefont
  {Acciarri} \emph {et~al.}}]{MicroBooNE:2016pwy}%
  \BibitemOpen
  \bibfield  {author} {\bibinfo {author} {\bibfnamefont {R.}~\bibnamefont
  {Acciarri}} \emph {et~al.} (\bibinfo {collaboration} {MicroBooNE}),\
  }\bibfield  {title} {\bibinfo {title} {{Design and Construction of the
  MicroBooNE Detector}},\ }\href
  {https://doi.org/10.1088/1748-0221/12/02/P02017} {\bibfield  {journal}
  {\bibinfo  {journal} {JINST}\ }\textbf {\bibinfo {volume} {12}}\bibfield
  {number} {\bibinfo  {number} { (02)},\ \bibinfo {pages} {P02017}},\ }\Eprint
  {https://arxiv.org/abs/1612.05824} {arXiv:1612.05824 [physics.ins-det]}
  \BibitemShut {NoStop}%
\bibitem [{\citenamefont {Acciarri}\ \emph {et~al.}(2018)\citenamefont
  {Acciarri} \emph {et~al.}}]{MicroBooNE:2017xvs}%
  \BibitemOpen
  \bibfield  {author} {\bibinfo {author} {\bibfnamefont {R.}~\bibnamefont
  {Acciarri}} \emph {et~al.} (\bibinfo {collaboration} {MicroBooNE}),\
  }\bibfield  {title} {\bibinfo {title} {{The Pandora multi-algorithm approach
  to automated pattern recognition of cosmic-ray muon and neutrino events in
  the MicroBooNE detector}},\ }\href
  {https://doi.org/10.1140/epjc/s10052-017-5481-6} {\bibfield  {journal}
  {\bibinfo  {journal} {Eur. Phys. J. C}\ }\textbf {\bibinfo {volume} {78}},\
  \bibinfo {pages} {82} (\bibinfo {year} {2018})},\ \Eprint
  {https://arxiv.org/abs/1708.03135} {arXiv:1708.03135 [hep-ex]} \BibitemShut
  {NoStop}%
\bibitem [{\citenamefont {Abi}\ \emph {et~al.}(2020{\natexlab{b}})\citenamefont
  {Abi} \emph {et~al.}}]{DUNE:2020ypp}%
  \BibitemOpen
  \bibfield  {author} {\bibinfo {author} {\bibfnamefont {B.}~\bibnamefont
  {Abi}} \emph {et~al.} (\bibinfo {collaboration} {DUNE}),\ }\href@noop {}
  {\bibinfo {title} {{Deep Underground Neutrino Experiment (DUNE), Far Detector
  Technical Design Report, Volume II: DUNE Physics}}} (\bibinfo {year}
  {2020}{\natexlab{b}}),\ \Eprint {https://arxiv.org/abs/2002.03005}
  {arXiv:2002.03005 [hep-ex]} \BibitemShut {NoStop}%
\bibitem [{Note1()}]{Note1}%
  \BibitemOpen
  \bibinfo {note} {\protect \texttt
  {https://github.com/vhewes/numl}}\BibitemShut {NoStop}%
\bibitem [{Note2()}]{Note2}%
  \BibitemOpen
  \bibinfo {note} {\protect \texttt
  {https://github.com/vhewes/pynuml}}\BibitemShut {NoStop}%
\bibitem [{\citenamefont {Misra}(2020)}]{misra2020mish}%
  \BibitemOpen
  \bibfield  {author} {\bibinfo {author} {\bibfnamefont {D.}~\bibnamefont
  {Misra}},\ }\href@noop {} {\bibinfo {title} {Mish: A self regularized
  non-monotonic activation function}} (\bibinfo {year} {2020}),\ \Eprint
  {https://arxiv.org/abs/1908.08681} {arXiv:1908.08681 [cs.LG]} \BibitemShut
  {NoStop}%
\bibitem [{\citenamefont {Krizhevsky}\ \emph {et~al.}(2017)\citenamefont
  {Krizhevsky}, \citenamefont {Sutskever},\ and\ \citenamefont
  {Hinton}}]{imagenet}%
  \BibitemOpen
  \bibfield  {author} {\bibinfo {author} {\bibfnamefont {A.}~\bibnamefont
  {Krizhevsky}}, \bibinfo {author} {\bibfnamefont {I.}~\bibnamefont
  {Sutskever}},\ and\ \bibinfo {author} {\bibfnamefont {G.~E.}\ \bibnamefont
  {Hinton}},\ }\bibfield  {title} {\bibinfo {title} {Imagenet classification
  with deep convolutional neural networks},\ }\href
  {https://doi.org/10.1145/3065386} {\bibfield  {journal} {\bibinfo  {journal}
  {Commun. ACM}\ }\textbf {\bibinfo {volume} {60}},\ \bibinfo {pages} {84–90}
  (\bibinfo {year} {2017})}\BibitemShut {NoStop}%
\bibitem [{\citenamefont {Paszke}\ \emph {et~al.}(2019)\citenamefont {Paszke},
  \citenamefont {Gross}, \citenamefont {Massa}, \citenamefont {Lerer},
  \citenamefont {Bradbury}, \citenamefont {Chanan}, \citenamefont {Killeen},
  \citenamefont {Lin}, \citenamefont {Gimelshein}, \citenamefont {Antiga},
  \citenamefont {Desmaison}, \citenamefont {Köpf}, \citenamefont {Yang},
  \citenamefont {DeVito}, \citenamefont {Raison}, \citenamefont {Tejani},
  \citenamefont {Chilamkurthy}, \citenamefont {Steiner}, \citenamefont {Fang},
  \citenamefont {Bai},\ and\ \citenamefont {Chintala}}]{paszke2019pytorch}%
  \BibitemOpen
  \bibfield  {author} {\bibinfo {author} {\bibfnamefont {A.}~\bibnamefont
  {Paszke}}, \bibinfo {author} {\bibfnamefont {S.}~\bibnamefont {Gross}},
  \bibinfo {author} {\bibfnamefont {F.}~\bibnamefont {Massa}}, \bibinfo
  {author} {\bibfnamefont {A.}~\bibnamefont {Lerer}}, \bibinfo {author}
  {\bibfnamefont {J.}~\bibnamefont {Bradbury}}, \bibinfo {author}
  {\bibfnamefont {G.}~\bibnamefont {Chanan}}, \bibinfo {author} {\bibfnamefont
  {T.}~\bibnamefont {Killeen}}, \bibinfo {author} {\bibfnamefont
  {Z.}~\bibnamefont {Lin}}, \bibinfo {author} {\bibfnamefont {N.}~\bibnamefont
  {Gimelshein}}, \bibinfo {author} {\bibfnamefont {L.}~\bibnamefont {Antiga}},
  \bibinfo {author} {\bibfnamefont {A.}~\bibnamefont {Desmaison}}, \bibinfo
  {author} {\bibfnamefont {A.}~\bibnamefont {Köpf}}, \bibinfo {author}
  {\bibfnamefont {E.}~\bibnamefont {Yang}}, \bibinfo {author} {\bibfnamefont
  {Z.}~\bibnamefont {DeVito}}, \bibinfo {author} {\bibfnamefont
  {M.}~\bibnamefont {Raison}}, \bibinfo {author} {\bibfnamefont
  {A.}~\bibnamefont {Tejani}}, \bibinfo {author} {\bibfnamefont
  {S.}~\bibnamefont {Chilamkurthy}}, \bibinfo {author} {\bibfnamefont
  {B.}~\bibnamefont {Steiner}}, \bibinfo {author} {\bibfnamefont
  {L.}~\bibnamefont {Fang}}, \bibinfo {author} {\bibfnamefont {J.}~\bibnamefont
  {Bai}},\ and\ \bibinfo {author} {\bibfnamefont {S.}~\bibnamefont
  {Chintala}},\ }\href@noop {} {\bibinfo {title} {Pytorch: An imperative style,
  high-performance deep learning library}} (\bibinfo {year} {2019}),\ \Eprint
  {https://arxiv.org/abs/1912.01703} {arXiv:1912.01703 [cs.LG]} \BibitemShut
  {NoStop}%
\bibitem [{\citenamefont {Fey}\ and\ \citenamefont
  {Lenssen}(2019)}]{fey2019fast}%
  \BibitemOpen
  \bibfield  {author} {\bibinfo {author} {\bibfnamefont {M.}~\bibnamefont
  {Fey}}\ and\ \bibinfo {author} {\bibfnamefont {J.~E.}\ \bibnamefont
  {Lenssen}},\ }\href@noop {} {\bibinfo {title} {Fast graph representation
  learning with pytorch geometric}} (\bibinfo {year} {2019}),\ \Eprint
  {https://arxiv.org/abs/1903.02428} {arXiv:1903.02428 [cs.LG]} \BibitemShut
  {NoStop}%
\bibitem [{\citenamefont {Loshchilov}\ and\ \citenamefont
  {Hutter}(2019)}]{loshchilov2019decoupled}%
  \BibitemOpen
  \bibfield  {author} {\bibinfo {author} {\bibfnamefont {I.}~\bibnamefont
  {Loshchilov}}\ and\ \bibinfo {author} {\bibfnamefont {F.}~\bibnamefont
  {Hutter}},\ }\href@noop {} {\bibinfo {title} {Decoupled weight decay
  regularization}} (\bibinfo {year} {2019}),\ \Eprint
  {https://arxiv.org/abs/1711.05101} {arXiv:1711.05101 [cs.LG]} \BibitemShut
  {NoStop}%
\bibitem [{\citenamefont {Smith}\ and\ \citenamefont
  {Topin}(2018)}]{smith2018superconvergence}%
  \BibitemOpen
  \bibfield  {author} {\bibinfo {author} {\bibfnamefont {L.~N.}\ \bibnamefont
  {Smith}}\ and\ \bibinfo {author} {\bibfnamefont {N.}~\bibnamefont {Topin}},\
  }\href@noop {} {\bibinfo {title} {Super-convergence: Very fast training of
  neural networks using large learning rates}} (\bibinfo {year} {2018}),\
  \Eprint {https://arxiv.org/abs/1708.07120} {arXiv:1708.07120 [cs.LG]}
  \BibitemShut {NoStop}%
\bibitem [{\citenamefont {Tian}\ \emph {et~al.}(2022)\citenamefont {Tian},
  \citenamefont {Mithun}, \citenamefont {Seymour}, \citenamefont {Chiu},\ and\
  \citenamefont {Kira}}]{tian2022striking}%
  \BibitemOpen
  \bibfield  {author} {\bibinfo {author} {\bibfnamefont {J.}~\bibnamefont
  {Tian}}, \bibinfo {author} {\bibfnamefont {N.}~\bibnamefont {Mithun}},
  \bibinfo {author} {\bibfnamefont {Z.}~\bibnamefont {Seymour}}, \bibinfo
  {author} {\bibfnamefont {H.-P.}\ \bibnamefont {Chiu}},\ and\ \bibinfo
  {author} {\bibfnamefont {Z.}~\bibnamefont {Kira}},\ }\href@noop {} {\bibinfo
  {title} {Striking the right balance: Recall loss for semantic segmentation}}
  (\bibinfo {year} {2022}),\ \Eprint {https://arxiv.org/abs/2106.14917}
  {arXiv:2106.14917 [cs.CV]} \BibitemShut {NoStop}%
\bibitem [{\citenamefont {Lin}\ \emph {et~al.}(2017)\citenamefont {Lin},
  \citenamefont {Goyal}, \citenamefont {Girshick}, \citenamefont {He},\ and\
  \citenamefont {Doll{\'a}r}}]{lin2017focal}%
  \BibitemOpen
  \bibfield  {author} {\bibinfo {author} {\bibfnamefont {T.-Y.}\ \bibnamefont
  {Lin}}, \bibinfo {author} {\bibfnamefont {P.}~\bibnamefont {Goyal}}, \bibinfo
  {author} {\bibfnamefont {R.}~\bibnamefont {Girshick}}, \bibinfo {author}
  {\bibfnamefont {K.}~\bibnamefont {He}},\ and\ \bibinfo {author}
  {\bibfnamefont {P.}~\bibnamefont {Doll{\'a}r}},\ }\bibfield  {title}
  {\bibinfo {title} {Focal loss for dense object detection},\ }in\ \href@noop
  {} {\emph {\bibinfo {booktitle} {Proceedings of the IEEE international
  conference on computer vision}}}\ (\bibinfo {year} {2017})\ pp.\ \bibinfo
  {pages} {2980--2988}\BibitemShut {NoStop}%
\bibitem [{\citenamefont {Kendall}\ \emph {et~al.}(2018)\citenamefont
  {Kendall}, \citenamefont {Gal},\ and\ \citenamefont
  {Cipolla}}]{kendall2018multitask}%
  \BibitemOpen
  \bibfield  {author} {\bibinfo {author} {\bibfnamefont {A.}~\bibnamefont
  {Kendall}}, \bibinfo {author} {\bibfnamefont {Y.}~\bibnamefont {Gal}},\ and\
  \bibinfo {author} {\bibfnamefont {R.}~\bibnamefont {Cipolla}},\ }\href@noop
  {} {\bibinfo {title} {Multi-task learning using uncertainty to weigh losses
  for scene geometry and semantics}} (\bibinfo {year} {2018}),\ \Eprint
  {https://arxiv.org/abs/1705.07115} {arXiv:1705.07115 [cs.CV]} \BibitemShut
  {NoStop}%
\bibitem [{\citenamefont {Snider}\ and\ \citenamefont
  {Petrillo}(2017)}]{Snider:2017wjd}%
  \BibitemOpen
  \bibfield  {author} {\bibinfo {author} {\bibfnamefont {E.~L.}\ \bibnamefont
  {Snider}}\ and\ \bibinfo {author} {\bibfnamefont {G.}~\bibnamefont
  {Petrillo}},\ }\bibfield  {title} {\bibinfo {title} {{LArSoft: Toolkit for
  Simulation, Reconstruction and Analysis of Liquid Argon TPC Neutrino
  Detectors}},\ }\href {https://doi.org/10.1088/1742-6596/898/4/042057}
  {\bibfield  {journal} {\bibinfo  {journal} {J. Phys. Conf. Ser.}\ }\textbf
  {\bibinfo {volume} {898}},\ \bibinfo {pages} {042057} (\bibinfo {year}
  {2017})}\BibitemShut {NoStop}%
\bibitem [{\citenamefont {Cai}\ \emph {et~al.}(2023)\citenamefont {Cai},
  \citenamefont {Herner}, \citenamefont {Yang}, \citenamefont {Wang},
  \citenamefont {Flechas}, \citenamefont {Harris}, \citenamefont {Holzman},
  \citenamefont {Pedro},\ and\ \citenamefont {Tran}}]{Cai:2023ldc}%
  \BibitemOpen
  \bibfield  {author} {\bibinfo {author} {\bibfnamefont {T.}~\bibnamefont
  {Cai}}, \bibinfo {author} {\bibfnamefont {K.}~\bibnamefont {Herner}},
  \bibinfo {author} {\bibfnamefont {T.}~\bibnamefont {Yang}}, \bibinfo {author}
  {\bibfnamefont {M.}~\bibnamefont {Wang}}, \bibinfo {author} {\bibfnamefont
  {M.~A.}\ \bibnamefont {Flechas}}, \bibinfo {author} {\bibfnamefont
  {P.}~\bibnamefont {Harris}}, \bibinfo {author} {\bibfnamefont
  {B.}~\bibnamefont {Holzman}}, \bibinfo {author} {\bibfnamefont
  {K.}~\bibnamefont {Pedro}},\ and\ \bibinfo {author} {\bibfnamefont
  {N.}~\bibnamefont {Tran}},\ }\bibfield  {title} {\bibinfo {title}
  {{Accelerating Machine Learning Inference with GPUs in ProtoDUNE Data
  Processing}},\ }\href {https://doi.org/10.1007/s41781-023-00101-0} {\bibfield
   {journal} {\bibinfo  {journal} {Comput. Softw. Big Sci.}\ }\textbf {\bibinfo
  {volume} {7}},\ \bibinfo {pages} {11} (\bibinfo {year} {2023})},\ \Eprint
  {https://arxiv.org/abs/2301.04633} {arXiv:2301.04633 [hep-ex]} \BibitemShut
  {NoStop}%
\bibitem [{\citenamefont {Abratenko}\ \emph {et~al.}(2023)\citenamefont
  {Abratenko} \emph {et~al.}}]{abratenko_2023_8370883}%
  \BibitemOpen
  \bibfield  {author} {\bibinfo {author} {\bibfnamefont {P.}~\bibnamefont
  {Abratenko}} \emph {et~al.} (\bibinfo {collaboration} {MicroBooNE}),\
  }\bibfield  {title} {\bibinfo {title} {{MicroBooNE BNB Inclusive Overlay
  Sample (No Wire Info)}},\ }\href {https://doi.org/10.5281/zenodo.8370883}
  {10.5281/zenodo.8370883} (\bibinfo {year} {2023})\BibitemShut {NoStop}%
\bibitem [{\citenamefont {Abratenko}\ \emph {et~al.}(2022)\citenamefont
  {Abratenko} \emph {et~al.}}]{abratenko_2022_7261921}%
  \BibitemOpen
  \bibfield  {author} {\bibinfo {author} {\bibfnamefont {P.}~\bibnamefont
  {Abratenko}} \emph {et~al.} (\bibinfo {collaboration} {MicroBooNE}),\
  }\bibfield  {title} {\bibinfo {title} {{MicroBooNE BNB Electron Neutrino
  Overlay Sample (No Wire Info)}},\ }\href
  {https://doi.org/10.5281/zenodo.7261921} {10.5281/zenodo.7261921} (\bibinfo
  {year} {2022})\BibitemShut {NoStop}%
\end{thebibliography}%

\end{document}